\documentclass[11pt,a4paper]{article}

\usepackage{fix-cm}                       
\usepackage{amsmath, amsthm, amssymb, amsfonts}  
\DeclareUnicodeCharacter{2212}{\ensuremath{-}}
\DeclareUnicodeCharacter{03B1}{\ensuremath{\alpha}}
\DeclareUnicodeCharacter{00B3}{\ensuremath{{}^3}}
\usepackage{graphicx}                     
\usepackage[table]{xcolor}                
\usepackage{hyperref}                     
\usepackage{caption}                      
\usepackage{array, tabularx}              
\usepackage{enumitem}                     
\usepackage{bm}                           
\usepackage[margin=1in]{geometry}

\usepackage[backend=biber, style=numeric, sorting=none]{biblatex}
\usepackage{xurl}                        
\usepackage{float}

\newcommand{\TableCellFormat}{%
  \raggedright
  \hyphenpenalty=10000
  \exhyphenpenalty=10000
  \sloppy
}
\newcolumntype{L}[1]{>{\TableCellFormat\arraybackslash}p{#1}}
\newcolumntype{Y}{>{\TableCellFormat\arraybackslash}X}
\newenvironment{tableitemize}
  {\begin{minipage}[t]{\linewidth}
   \TableCellFormat
   \vspace{0.15\baselineskip}
   \begin{itemize}[leftmargin=1.2em, rightmargin=0.6em, labelsep=0.45em, itemsep=0.35\baselineskip, parsep=0pt, topsep=0.1\baselineskip, partopsep=0pt]}
  {\end{itemize}
   \vspace{0.45\baselineskip}
   \end{minipage}}

\numberwithin{equation}{section}

\newtheorem{theorem}{Theorem}[section]
\newtheorem{lemma}[theorem]{Lemma}

\begin{document}

\title{Generating integrable systems from an autonomous ordinary differential equation}
\author{Graham Hesketh \\
\small \texttt{gdh1e10@gmail.com}}
\date{October 19, 2026}
\maketitle

\begin{abstract}
The freedom to realise special-function dynamics within integrable Kummer-type Hamiltonian systems remains underexplored.
We investigate this freedom through examples constructed from prescribed autonomous scalar equations, using the established reduction and mode-reconstruction framework.
A local holomorphic equation $\rho'=f(\rho)$ determines a family of canonical Hamiltonians on a selected energy level, subject to regularity and compatible branch choices.
We introduce the category of log-augmented polynomial systems, whose polynomial mode equations retain logarithmic terms in their scalar reductions, and illustrate it with examples including the Lotka--Volterra model and a parity-time-symmetric dimer.
For rational generating equations, we prove a correspondence with extended Lambert--W-type inverse laws and establish elementary evaluation of rational reconstruction quadratures, giving local solvability results for the associated coupled systems.
Further constructions use inverse incomplete beta and inverse spherical Bessel functions and the modular $\lambda$ function.
The modular example has explicit theta-constant mode solutions and a direct map to the Darboux--Halphen system.
\end{abstract}


\section{Introduction}

Integrable coupled-mode systems connect nonlinear optics with special-function theory. Resonant wave mixing and nonlinear directional couplers provide familiar examples \cite{armstrongInteractionsLightWaves1962,stolenParametricAmplificationFrequency1982,chenFourphotonParametricMixing1989,chenFourwaveMixingOptical1989,agrawalNonlinearFiberOptics2019,jensenNonlinearCoherentCoupler1982}, whose conserved modal-power combinations reduce the dynamics to a scalar autonomous differential equation. Cubic and quartic polynomial reductions lead to elliptic functions, with the individual mode amplitudes recovered by further quadratures \cite{jensenNonlinearCoherentCoupler1982,heskethGeneralComplexEnvelope2015,heskethFWMWeierstrassElliptic2026,heskethCompleteWeierstrassElliptic2026}. Under suitable parameter choices and initial conditions, these systems also connect with classical mechanics through projections to the Lagrange top and Duffing oscillator, while the Euler top appears as a special reduction \cite{Whittaker1937,kovacicJacobiEllipticFunctions2016}. More recently, we obtained local solutions for polynomial Kummer-type systems with interaction terms formed from products of $N>4$ modes, using hyperelliptic sigma functions and Abelian integrals \cite{zotero-item-855}.

The construction of integrable Hamiltonian systems has several established approaches. Lax representations relate the equations of motion to spectral invariants, while separation relations can be used to define commuting Hamiltonians and their separated equations. Stäckel transforms and coupling-constant metamorphosis provide ways to obtain related integrable systems from known ones \cite{krishnaswamiIntroductionLaxPairs2020,blaszakGeneralizedStackelSystems2011,sergyeyevCouplingConstantMetamorphosis2012}. These approaches offer a broad context for the present work, which concerns a particular family of resonantly coupled modes and uses a scalar autonomous equation as the starting point for its construction.

The closest precedents are the resonant Hamiltonians studied by Kummer and their subsequent generalisations. Kummer developed reduction and integration for Hamiltonians with polynomial dependence on oscillator powers and constant resonant coupling \cite{kummerResonantClassicalHamiltonians1990}. Related work on integrable multiboson systems and intensity-dependent conversion allowed functions of mode occupations to enter both the diagonal and interaction terms, with explicit constructions linked to orthogonal polynomials \cite{odzijewiczIntegrableMultibosonSystems2001,horowskiSystemsIntensitydependentConversion2004}. Odzijewicz and Wawreniuk formulated a classical and quantum Kummer-shape framework with smooth power-dependent functions in place of the polynomial and constant choices. Their examples include a non-polynomial classical Hamiltonian associated with a deformed oscillator algebra \cite{odzijewiczClassicalQuantumKummer2016}. Their studies of three- and four-wave mixing further connect this framework with nonlinear optics \cite{odzijewiczIntegrabilityCorrespondenceClassical2018,odzijewiczIntegrableClassicalQuantum2020}.

We believe that the freedom to realise a wider range of special functions in Kummer-type systems remains underexplored. The established freedom in the coefficient functions makes it possible to explore scalar dynamics beyond the familiar elliptic examples. Polynomial mode equations can themselves conceal such behaviour: logarithmic terms in a conserved Hamiltonian may disappear from the equations of motion after the conserved value is used, yet remain in the reduced scalar equation. The classical two-species Lotka--Volterra predator--prey system is a familiar example: its polynomial equations possess a logarithmic first integral, and the scalar reduction considered below retains logarithmic terms \cite{baigentLotkaVolterraDynamicsIntroduction}. Polynomial coupled equations therefore need not lead to an elliptic or hyperelliptic reduction. Conversely, allowing non-polynomial Hamiltonians provides a way to embed prescribed scalar dynamics and special functions into coupled canonical systems. These observations motivate treating the scalar equation as a means both of constructing systems and of comparing their dynamics.

\subsection{Our contribution}
\label{sec:our-contribution}

{\emergencystretch=1em
We explore the special-function freedom of the established Kummer-type Hamiltonian framework through explicit examples and solvability results. Our approach starts from a prescribed scalar autonomous equation and constructs compatible canonical systems, with individual modes recovered by quadratures.\par}

First, we introduce the category of \emph{log-augmented polynomial systems}, whose polynomial mode equations retain logarithmic terms in their scalar reductions. We connect examples including Lotka--Volterra and a parity-time-symmetric dimer with newly constructed mode systems, showing that polynomial coupled equations need not have algebraic scalar dynamics.

Second, we prove a correspondence between rational generating equations and extended Lambert--$W$-type inverse laws. We also establish elementary evaluation of rational reconstruction quadratures and describe how the associated rational coupled systems can be converted locally into polynomial systems by reparameterisation.

Third, we give constructions involving inverse incomplete beta and inverse spherical Bessel functions, together with a system generated by the modular $\lambda$ function. The modular example has explicit theta-constant mode solutions and a direct map to the Darboux--Halphen system. A formulation in terms of total modal phase explains how prescribed phases can be incorporated, with a correction by quadrature when required.

These contributions use the known reduction and reconstruction mechanism to investigate a broader range of special-function dynamics.

\subsection{Organisation of the paper}
\label{sec:organisation}

We first recall the general Hamiltonian reduction and develop the inverse construction. We then review elliptic and higher-degree examples, introduce log-augmented systems, and establish the rational generating-equation results. The modular $\lambda$ construction follows. The final technical section develops the total-phase formulation and illustrates it with the inverse spherical Bessel example. Supporting integrability and Weierstrass-reparameterisation arguments are supplied in supplementary notebooks accompanying this paper.

\section{The general Hamiltonian system}
\label{sec:gen-ham-system}
In this section we recall the established Hamiltonian reduction and reconstruction mechanism in the notation used for the subsequent construction.

The separation and reconstruction mechanism is established in the Kummer literature. Kummer treated constant resonant coupling ($P$ constant in our notation) and polynomial power dependence ($Q$ polynomial), subject to the degree restrictions of his setting \cite{kummerResonantClassicalHamiltonians1990}. Odzijewicz and Wawreniuk allowed smooth power-dependent coupling and diagonal functions, corresponding to $P$ and $Q$, and obtained the separated scalar equation and reconstruction quadratures \cite[Section~2, especially Eq.~(2.31)]{odzijewiczClassicalQuantumKummer2016}. Their formulation uses conjugate oscillator variables and a positive coupling function; here we work locally with independent complex canonical variables and holomorphic coefficients.

The notation is motivated by nonlinear optics and elliptic function theory.
Unless otherwise stated, the complex constructions are local: $P,Q$ are holomorphic on an open domain $\Omega\subset\mathbb C^N$, and the modes are holomorphic on a simply connected open domain $D\subset\mathbb C$ with $x(D)\subset\Omega$.
Meromorphic coefficients are restricted to domains avoiding their poles; roots, logarithms, and inverse functions are taken on compatible local branches, and every displayed division is restricted to a nonzero denominator.
Real restrictions may be taken where these expressions are defined. These local statements assert neither global single-valuedness nor continuation through singularities.

Let $N\in\mathbb Z^+$ and consider canonical pairs $u_j,v_j:D\to\mathbb C$ $(j=1,\ldots,N)$, with $v_j(z)$ the Hamiltonian conjugate of $u_j(z)$.
In analogy with nonlinear optics we refer to $u_j, v_j$ as modes and the product $u_j v_j$ as modal power. 
The modes are functions of a single independent variable $z$ that might represent propagation length in nonlinear optics or time in quantum mechanics. 
For $P,Q:\Omega\to\mathbb C$ as above and integers $\beta_m\in\mathbb Z$ $(m=1,\ldots,N)$ with $\sum_m\beta_m\ne0$, define the canonical Hamiltonian:
\begin{equation}
    H\left(u_1,\dots, u_N, v_1,\dots, v_N\right) = \left(\prod_{m=1}^{N} u_{m}^{{\beta}_{m}} + \prod_{m=1}^{N} v_{m}^{{\beta}_{m}}\right) P{\left(u_1 v_1, \dots , u_N v_N \right)} - Q{\left(u_1 v_1, \dots , u_N v_N\right)} \label{eq:Hamiltonian}
\end{equation}
where if $\beta_j>1$ it indicates degeneracy in the system such as that which appears in degenerate four-wave mixing in nonlinear optics.
Once we have obtained the equations of motion from the Hamiltonian we then work on the zero energy level $H=0$. 
This is not restrictive since any constant energy value can be absorbed into $Q$ as an additive constant. 
Thus we have:
\begin{equation}
 \left(\prod_{m=1}^{N} u_{m}^{{\beta}_{m}} + \prod_{m=1}^{N} v_{m}^{{\beta}_{m}}\right) P = Q. \label{eq:u-v-prod-P-Q-pos}
\end{equation} 
Let us introduce the notation $x_j = u_j v_j$ such that:
\begin{align}
    P{\left(u_1 v_1, \dots , u_N v_N \right)} &= P{\left(x_1, \dots , x_N \right)}, \\
    \partial_{v_j} P &= u_j \partial_{x_j} P = u_j \left(\partial_{x_j} P\right) |_{x_j=u_j v_j},
\end{align}
then we can write the equations of motion obtained via Hamilton's equations as:
\begin{align}
    \frac{d u_j}{dz} &= \partial_{v_j} H, \notag \\
        &= \left\{\left(\prod_{m=1}^{N} u_{m}^{{\beta}_{m}} + \prod_{m=1}^{N} v_{m}^{{\beta}_{m}}\right) \partial_{x_j} P - \partial_{x_j} Q \right\} u_j + \frac{\beta_j P}{v_j} \prod_{m=1}^{N} v_{m}^{{\beta}_{m}}, \notag \\
        &= \left\{\frac{Q \partial_{x_j} P}{P} - \partial_{x_j} Q \right\} u_j + \frac{\beta_j P}{v_j} \prod_{m=1}^{N} v_{m}^{{\beta}_{m}}, \notag \\[10pt]
    \frac{d v_j}{dz} &= -\partial_{u_j} H, \notag \\
        &= -\left\{\left(\prod_{m=1}^{N} u_{m}^{{\beta}_{m}} + \prod_{m=1}^{N} v_{m}^{{\beta}_{m}}\right) \partial_{x_j} P - \partial_{x_j} Q \right\} v_j - \frac{\beta_j P}{u_j} \prod_{m=1}^{N} u_{m}^{{\beta}_{m}}, \notag \\
        &= -\left\{\frac{Q \partial_{x_j} P}{P} - \partial_{x_j} Q \right\} v_j - \frac{\beta_j P}{u_j} \prod_{m=1}^{N} u_{m}^{{\beta}_{m}}, \label{eq:eoms}
\end{align}
where the terms inside the curly brackets depend only on the modal powers $u_j v_j$ and thus, in analogy with nonlinear optics, could be viewed as phase modulation effects, whereas the final product terms deliver wave mixing among modes. 
It follows from \eqref{eq:eoms} that modal power evolves according to:
\begin{equation}
    \frac{d \left(u_j v_j\right)}{dz} = - \left(\prod_{m=1}^{N} u_{m}^{{\beta}_{m}} - \prod_{m=1}^{N} v_{m}^{{\beta}_{m}}\right) \beta_j P, \label{eq:duvj-u-v-prod-minus}
\end{equation}
and thus we have intermodal power conservation laws, which express the conserved weighted power combinations known in optics as Manley--Rowe relations \cite{armstrongInteractionsLightWaves1962}:
\begin{equation}
    \beta_k \frac{d \left(u_j v_j\right)}{dz} - \beta_j \frac{d \left(u_k v_k\right)}{dz}  = 0. \label{eq:manley-rowe-diff-beta}
\end{equation}

Let us now define the function $\rho(z) : \mathbb{C} \to \mathbb{C}$ as the negative total modal power over the sum of the degeneracy numbers $\beta_j$, which if all $\beta_j=1$ would be the negative mean modal power:
\begin{align}
    \rho(z) &= - \frac{\sum_{j=1}^{N}u_j v_j}{\sum_{j=1}^{N}\beta_j}, \label{eq:rho-mean-mode-pow}
\end{align}
The assumption $\sum_j\beta_j\ne0$ makes this normalization well defined. It follows that:
\begin{align}
    \rho' &= \frac{d \rho}{dz}  = \left(\prod_{m=1}^{N} u_{m}^{{\beta}_{m}} - \prod_{m=1}^{N} v_{m}^{{\beta}_{m}}\right) P, \label{eq:rho-prime-u-v-prod} \\
    \frac{d \left(u_j v_j\right)}{dz} &= -\beta_j \rho', \label{eq:duvj-rho} \\
    u_j v_j &= \gamma_j - \beta_j \rho, \label{eq:uvj-rho-gammaj} \\
    \sum_{j=1}^{N}\gamma_j &= 0
\end{align}
where the integration constants $\gamma_j$ sum to zero by the definition of $\rho(z)$ in \eqref{eq:rho-mean-mode-pow}. 
We thus see that all modal powers can be expressed in terms of a single $z$-dependent function $\rho(z)$ and therefore $Q$ and $P$ can be expressed as functions of $\rho(z)$.
Combining \eqref{eq:u-v-prod-P-Q-pos}, \eqref{eq:duvj-u-v-prod-minus}, and \eqref{eq:duvj-rho} we obtain the important relations:
\begin{align}
    \prod_{m=1}^{N} u_{m}^{{\beta}_{m}} &= \frac{Q + \rho'}{2P}, \notag \\
    \prod_{m=1}^{N} v_{m}^{{\beta}_{m}} &= \frac{Q - \rho'}{2P}. \label{eq:-u-v-prod-fs-det}
\end{align}
Section \ref{sec:elliptic-examples} places several nonlinear-optical models within this framework.
In our previous work on four-wave mixing and coherent couplers, we connected the product relations \eqref{eq:-u-v-prod-fs-det} with the Frobenius--Stickelberger determinant identity for elliptic functions \cite{heskethFWMWeierstrassElliptic2026,heskethCompleteWeierstrassElliptic2026}.
In the companion work on hyperelliptic extensions, we associate the corresponding relations with \^Onishi's hyperelliptic determinant identities and give them a Schur-complement interpretation \cite{zotero-item-855,onishiDETERMINANTEXPRESSIONSHYPERELLIPTIC2005}.
These connections suggest that the product relations encode structure beyond their immediate role in reducing the equations; the detailed determinant constructions are developed in those works.
Here we use \eqref{eq:-u-v-prod-fs-det} to recast the equations of motion in \eqref{eq:eoms} as:
\begin{align}
    \frac{d \log{u_j}}{dz} &= \left\{\frac{Q \partial_{x_j} P}{P} - \partial_{x_j} Q \right\} + \frac{\beta_j \left(Q - \rho'\right)}{2 (\gamma_j - \beta_j \rho)}, \notag \\[10pt]
    \frac{d \log{v_j}}{dz} &= -\left\{\frac{Q \partial_{x_j} P}{P} - \partial_{x_j} Q \right\} - \frac{\beta_j \left(Q + \rho'\right)}{2 (\gamma_j - \beta_j \rho)}, \label{eq:eoms-rho-only}
\end{align}
thus demonstrating that the logarithmic derivatives of $u_j, v_j$ are separable and can be expressed in terms of the chosen $Q, P$ and $\rho, \rho'$.

We now return to the defining equation for $\rho'$. Observe the simple but important algebraic identity:
\begin{equation}
    \left(\prod_{m=1}^{N} u_{m}^{{\beta}_{m}} + \prod_{m=1}^{N} v_{m}^{{\beta}_{m}}\right)^2 -  \left(\prod_{m=1}^{N} u_{m}^{{\beta}_{m}} - \prod_{m=1}^{N} v_{m}^{{\beta}_{m}}\right)^2 = 4\prod_{m=1}^{N} \left(u_{m}v_m\right)^{{\beta}_{m}} \label{eq:-uv-prod-sqr-id}
\end{equation}
This important identity is what delivers separability for the $\rho'$ equation as we note that the equation for $\rho'$ contains the difference of products over $u_j$ and $v_j$ and the Hamiltonian contains their sum. 
It thus follows either by squaring \eqref{eq:rho-prime-u-v-prod} and using \eqref{eq:u-v-prod-P-Q-pos} and \eqref{eq:-uv-prod-sqr-id}, or equivalently, from multiplying the two expressions in \eqref{eq:-u-v-prod-fs-det}, that:
\begin{align}
    \label{eq:drho-sqrd-ham}
    {\rho'}^2 &= Q^2 - 4P^2 \prod_{m=1}^{N} \left(u_{m}v_m\right)^{{\beta}_{m}}, \notag \\
    {\rho'}^2 &= Q^2 - 4P^2 \prod_{m=1}^{N} \left(\gamma_m - \beta_m \rho\right)^{{\beta}_{m}},
\end{align}
and thus the differential equation for $\rho$ is separable. 

Choosing $\beta_N\ne0$ after relabelling if necessary, the $N-1$ quantities
\begin{equation}
    C_j=\beta_Nu_jv_j-\beta_ju_Nv_N,\qquad j=1,\ldots,N-1,
\end{equation}
commute with one another and with $H$. The system is therefore Liouville integrable on the regular locus where these $N$ integrals are functionally independent. A detailed verification is supplied in the supplementary computational notebooks. The constructions below use the explicit scalar reduction and reconstruction quadratures directly. No global invariant-torus or action--angle conclusion is asserted for the general complex systems.

An independent differential-equation proof of the reconstruction identities is given in Appendix~\ref{app:direct-reconstruction}. Reconstruction requires compatible initial products; constant modal powers require an additional compatibility condition, also treated there.

\section{Generating integrable systems from an autonomous differential equation}
\label{sec:ode-generation}

We have shown in Section \ref{sec:gen-ham-system} that we may pick $Q, P$ and obtain from them an integrable system in which the equations of motion for $u_j, v_j, \rho$ are separable.
In some cases we may know the $Q,P$ from which we wish to start, for example we may be working with elliptic functions and we may recognise or choose a Hamiltonian as some rearrangement of the Frobenius-Stickelberger determinant.
However, we may on other occasions wish to generate a new integrable system from some known autonomous differential equation and we may not know $Q, P$ at the outset.
In this section we show how we may get $Q, P$ quite simply from an autonomous differential equation, thereby generating an integrable system from it.

Let $f$ be holomorphic on an open set $\mathcal U\subset\mathbb C$, and let $\rho:D\to\mathcal U$ be a local holomorphic solution of the autonomous equation
\begin{equation}
    \rho' = f(\rho)
\end{equation}
Using $x_j=u_j v_j=\gamma_j-\beta_j\rho$ from \eqref{eq:uvj-rho-gammaj}, we write:
\begin{equation}
    f(\rho) = F(x_1,\dots,x_N). \label{eq:f-F}
\end{equation}
The embedding of $\rho$ into the $x_j$ in \eqref{eq:f-F} may manifest in two distinct ways. 
Firstly, it may be that $F$ treats each $x_j$ differently, for example if $f(\rho)$ was a polynomial in $\rho$, 
$F$ may be a polynomial that raises each $x_j$ to different powers but still upon substitution of the $x_j$ for $\rho$ produces the correct coefficients from $f$, 
this is common in physical nonlinear optics cases with elliptic function solutions for example.
Alternatively, we may simply substitute the sum over $u_j v_j$ from \eqref{eq:rho-mean-mode-pow} into $f$ to define $F$, 
which is a simple approach for symmetric toy model exploration. 
Writing $p=P(x(\rho))$, $q=Q(x(\rho))$ and $r=\prod_jx_j(\rho)^{\beta_j}$, equation \eqref{eq:drho-sqrd-ham} gives:
\begin{align}
    {\rho'}^2 &= q^2 - 4p^2r, \\
    f(\rho)^2 &= q^2 - 4p^2r,
\end{align}
and for consistency set:
\begin{equation}
    F(x_1,\dots,x_N)^2 = Q(x_1,\dots,x_N)^2 - 4P(x_1,\dots,x_N)^2R(x_1,\dots,x_N)
\end{equation}
with:
\begin{equation}
    R(x_1,\dots,x_N) = \prod_{m=1}^{N}x_m^{\beta_m}.
\end{equation}
We may solve for $Q$ with $\varsigma=\pm1$:
\begin{equation}
    Q = \varsigma \sqrt{F^2 + 4P^2R}, \label{eq:Q-F-P-R}
\end{equation}
and use this to replace $Q$ in \eqref{eq:Hamiltonian} yielding the Hamiltonian:
\begin{multline}
    H\left(u_1,\dots, u_N, v_1,\dots, v_N\right) = \\
    - \varsigma \sqrt{F{\left(u_1 v_1, \dots , u_N v_N\right)}^2 + 4 P{\left(u_1 v_1, \dots , u_N v_N\right)}^2R{\left(u_1 v_1, \dots , u_N v_N\right)}} \\
    + P{\left(u_1 v_1, \dots , u_N v_N \right)}\left(\prod_{m=1}^{N} u_{m}^{{\beta}_{m}} + \prod_{m=1}^{N} v_{m}^{{\beta}_{m}}\right).
\end{multline}
The equations of motion can again be derived using \eqref{eq:eoms} where one would substitute \eqref{eq:Q-F-P-R} and note:
\begin{align}
    \partial_{x_j}Q &= \varsigma \partial_{x_j} \sqrt{F^2 + 4P^2R}, \notag \\
    &= \frac{\varsigma}{2\sqrt{F^2 + 4P^2R}} \left(2F\partial_{x_j}F + 8PR\partial_{x_j}P + 4P^2\partial_{x_j}R\right), \notag \\
    &= \frac{\varsigma}{\sqrt{F^2 + 4P^2R}} \left(F\partial_{x_j}F + 4PR\partial_{x_j}P + \frac{2\beta_j P^2 R}{x_j}\right).
\end{align}
where, having obtained the equations of motion we again work on the zero energy level such that $H=0$ and \eqref{eq:u-v-prod-P-Q-pos} still holds. 
The same calculation as in Section \ref{sec:gen-ham-system} shows that the system again conserves the Hamiltonian and the Manley-Rowe like intermodal power laws that stem from \eqref{eq:manley-rowe-diff-beta}. 
Furthermore, the differential equations for $u_j, v_j$ indeed yield the target:
\begin{equation}
    {\rho'}^2=f(\rho)^2,
\end{equation} 
where the branch of $\rho'=\pm f(\rho)$ is fixed by the initial condition. 
We note that although $Q$ is substituted in terms of $F$, one may still choose $P$ subject to the local regularity conditions, so $f$ does not generate a unique system.
In particular, $F$ must have a holomorphic extension to the modal-power domain, and $F^2+4P^2R$ must admit the selected holomorphic square root defining $Q$. A nonzero radicand at the initial point is sufficient after shrinking the domain.
The reconstruction formulas require $P$ and the modal products to be nonzero; the direct reconstruction theorem in Appendix~\ref{app:direct-reconstruction} applies where $f(\rho)\ne0$, with stationary cases treated separately there.

\begin{lemma}[Removal of a common modal-power factor]
\label{lem:common-modal-power-factor}
Consider the Hamiltonian in \eqref{eq:Hamiltonian} on a region where $P(x_1,\dots,x_N)\ne0$, and define
\begin{equation}
    \widehat{Q}(x_1,\dots,x_N)
    =\frac{Q(x_1,\dots,x_N)}{P(x_1,\dots,x_N)}.
\end{equation}
Then
\begin{equation}
    H=P\widehat{H},
    \qquad
    \widehat{H}
    =\prod_{m=1}^{N}u_m^{\beta_m}
    +\prod_{m=1}^{N}v_m^{\beta_m}
    -\widehat{Q}(x_1,\dots,x_N).
    \label{eq:Hhat-time-reparam}
\end{equation}
On the level set $H=0$, the solutions generated by $H$ and $\widehat{H}$ describe the same curves in the variables $u_j,v_j$ after a local reparameterisation of the independent variable.
\end{lemma}

\begin{proof}
Since $P\ne0$, the level sets $H=0$ and $\widehat{H}=0$ coincide. Hamilton's equations and the product rule give
\begin{align}
    \frac{du_j}{dz}
    &=P\frac{\partial\widehat{H}}{\partial v_j}
    +\widehat{H}\frac{\partial P}{\partial v_j},
    \notag\\
    \frac{dv_j}{dz}
    &=-P\frac{\partial\widehat{H}}{\partial u_j}
    -\widehat{H}\frac{\partial P}{\partial u_j}.
\end{align}
Consequently, on $\widehat{H}=0$,
\begin{equation}
    \frac{du_j}{dz}
    =P\frac{\partial\widehat{H}}{\partial v_j},
    \qquad
    \frac{dv_j}{dz}
    =-P\frac{\partial\widehat{H}}{\partial u_j}.
    \label{eq:H-Hhat-vector-fields}
\end{equation}
Define a new local parameter $\xi$ along a solution by
\begin{equation}
    \frac{d\xi}{dz}
    =P(x_1(z),\dots,x_N(z)).
    \label{eq:xi-z-P-reparam}
\end{equation}
If $\widehat{u}_j(\xi)=u_j(z(\xi))$, $\widehat{v}_j(\xi)=v_j(z(\xi))$, and $\widehat{x}_j=\widehat{u}_j\widehat{v}_j$, then \eqref{eq:H-Hhat-vector-fields} becomes
\begin{equation}
    \frac{d\widehat{u}_j}{d\xi}
    =\frac{\partial\widehat{H}}{\partial\widehat{v}_j},
    \qquad
    \frac{d\widehat{v}_j}{d\xi}
    =-\frac{\partial\widehat{H}}{\partial\widehat{u}_j}.
\end{equation}
Equivalently, the inverse change of parameter is determined by
\begin{equation}
    z(\xi)
    =z_0+
    \int_{\xi_0}^{\xi}
    \frac{d\eta}
    {P(\widehat{x}_1(\eta),\dots,\widehat{x}_N(\eta))}.
    \label{eq:z-xi-P-reparam}
\end{equation}
Thus $\widehat{H}=H/P$ generates the same solution curves on the zero-energy level with the common factor $P$ removed from the parametrisation.
\end{proof}

The reparameterised mode equations are explicitly
\begin{align}
    \frac{d\widehat{u}_j}{d\xi}
    &=-\widehat{u}_j\partial_{\widehat{x}_j}\widehat{Q}
    +\frac{\beta_j}{\widehat{v}_j}
    \prod_{m=1}^{N}\widehat{v}_m^{\beta_m},
    \notag\\
    \frac{d\widehat{v}_j}{d\xi}
    &=\widehat{v}_j\partial_{\widehat{x}_j}\widehat{Q}
    -\frac{\beta_j}{\widehat{u}_j}
    \prod_{m=1}^{N}\widehat{u}_m^{\beta_m}.
    \label{eq:Hhat-reparameterised-modes}
\end{align}
Define $\widehat{\rho}(\xi)=\rho(z(\xi))$. Then $\widehat{x}_j=\gamma_j-\beta_j\widehat{\rho}$, and the associated scalar equation is
\begin{equation}
    \left(\frac{d\widehat{\rho}}{d\xi}\right)^2
    =\widehat{Q}^{\,2}
    -4\prod_{m=1}^{N}
    \left(\gamma_m-\beta_m\widehat{\rho}\right)^{\beta_m}.
    \label{eq:Hhat-reparameterised-rho}
\end{equation}
Thus the factorisation is always available locally by taking $\widehat{Q}=Q/P$, but it does not in general make the original choice of $P$ irrelevant: $P$ may remain within the ratio $Q/P$ and it fixes the relation between $z$ and $\xi$ through \eqref{eq:xi-z-P-reparam}.
The result is most useful when $P$ and $Q$ have an explicit common factor which cancels from $\widehat{Q}$, in which case that factor affects only the parametrisation on the chosen energy level.
The modular $\lambda$ system in Section \ref{sec:modular-example} provides such an example, since the common factor involving the complete elliptic integral can be transferred into the change of parameter.
For broader discussion of such time reparameterisations, see \cite{garcia-naranjoStructurePreservingDiscretization2020}.
The same argument applies on any fixed energy level after subtracting the energy from $H$, as is implemented here by an additive change to $Q$.

\subsection{Recognising equivalent Hamiltonian forms}
\label{sec:equivalent-mixing-functions}

A Hamiltonian that depends nonlinearly on the mixing terms may also admit a representation within the family considered here on a chosen level set.
To see this, write $x=(x_1,\ldots,x_N)$, $U=\prod_j u_j^{\beta_j}$, $V=\prod_j v_j^{\beta_j}$, and consider
\begin{equation}
    \mathcal H_S=P(x)S(U+V)-Q(x).
\end{equation}
Work locally where $P\ne0$ and $S$ admits a differentiable local inverse with nonvanishing derivative; in the complex setting, take a regular holomorphic inverse branch.
On the corresponding branch of $\mathcal H_S=0$, applying $S^{-1}$ gives
\begin{equation}
    U+V=\widetilde Q(x),
    \qquad \widetilde Q(x)=S^{-1}\!\left(\frac{Q(x)}{P(x)}\right).
\end{equation}
Thus the same local level set is described by $K=U+V-\widetilde Q(x)=0$, a Hamiltonian of the original form with coupling function $1$.
Equality of the level sets alone does not identify their parametrised dynamics.
Indeed, differentiating $S(\widetilde Q)=Q/P$ shows that, on this level, Hamilton's equations take the form
\begin{equation}
    \frac{du_j}{dz}=\partial_{v_j}\mathcal H_S=A(x)\partial_{v_j}K,
    \qquad
    \frac{dv_j}{dz}=-\partial_{u_j}\mathcal H_S=-A(x)\partial_{u_j}K,
    \label{eq:nonlinear-mixing-vector-factor}
\end{equation}
where $A(x)=P(x)S'(\widetilde Q(x))$. Thus the mode derivatives differ from those generated by $K$ by the same nonzero factor $A(x)$. The solution curves therefore agree after the local change of parameter $d\xi/dz=A(x(z))$, as in Lemma \ref{lem:common-modal-power-factor}.
Alternatively, define
\begin{equation}
    \widetilde H=A(x)K
    =A(x)(U+V)-A(x)\widetilde Q(x).
\end{equation}
This has the original Hamiltonian form with coefficients $A$ and $A\widetilde Q$. On $K=0$, the terms produced by differentiating $A$ in $\widetilde H=AK$ vanish, so this Hamiltonian gives the same Hamilton equations as $\mathcal H_S$ on the selected level in the original variable $z$.
Consequently, a locally invertible function of $U+V$ does not enlarge the class of level-set dynamics represented by arbitrary $P,Q$, although this equivalence may not be apparent from the initial Hamiltonian expression.
The equivalence is local to the chosen inverse branch and level set; the Hamiltonians need not generate the same dynamics away from that level.

\section{Nonlinear optics, classical mechanics, and elliptic functions}
\label{sec:elliptic-examples}

Continuous-wave and quasi-continuous-wave nonlinear optics, where temporal derivatives can be neglected, provides a broad hierarchy of integrable coupled-mode systems, many with elliptic-function solutions. Here we recap the importance and extent of this hierarchy by presenting a common system and summarising its optical and classical-mechanical special cases. We then show how the same framework extends beyond the elliptic hierarchy to higher-degree scalar curves and incomplete beta inversion. Detailed solutions of the established optical models are left to the cited literature.

For constants $a_j, a_{j,k}, c \in \mathbb{C}$, let us consider the $N \le 4$ mode system with:
\begin{align}
\label{eq:Q-elliptic-system}
Q\left(x_1,\dots,x_N\right) &= {a}_{0} + \sum_{j=1}^{N} {a}_{j} {x}_{j} + \frac{1}{2}\sum_{j=1, k=1}^{N} {a}_{j,k} {x}_{j} {x}_{k}, \notag \\
P\left(x_1,\dots,x_N\right) &= c
\end{align}
such that by \eqref{eq:Hamiltonian} the Hamiltonian is:
\begin{equation}
    H\left(u_1,\dots, u_N, v_1,\dots, v_N\right) = c\left(\prod_{m=1}^{N} u_{m}^{{\beta}_{m}} + \prod_{m=1}^{N} v_{m}^{{\beta}_{m}}\right) - {a}_{0} - \sum_{j=1}^{N} {a}_{j} u_j v_j - \frac{1}{2}\sum_{j=1, k=1}^{N} {a}_{j,k} u_j v_j u_k v_k
\end{equation} 
with $a_{j,k}=a_{k,j}$, and Hamilton's equations give:
\begin{align}
\frac{\partial}{\partial z} u_j &= \partial_{v_j}H = -\left({a}_{j} + \sum_{k=1}^{N} {a}_{j,k} u_k v_k \right) u_j + \frac{c {\beta}_{j}}{v_j} \prod_{k=1}^{N} v^{{\beta}_{k}}_k, \notag \\
\frac{\partial}{\partial z} v_j &= -\partial_{u_j}H =\left({a}_{j} + \sum_{k=1}^{N} {a}_{j,k} u_k v_k \right) v_j - \frac{c {\beta}_{j}}{u_j} \prod_{k=1}^{N} u^{{\beta}_{k}}_k. \label{eq:elliptic-system}
\end{align}
As before, after deriving the equations of motion we restrict to the zero-energy level $H=0$; the constant $a_0$ offsets this energy level within $Q$.

By \eqref{eq:uvj-rho-gammaj} and \eqref{eq:drho-sqrd-ham} we have:
\begin{equation}
    \label{eq:drho-sqrd-elliptic-examples}
    \rho'^{2} = \left({a}_{0} + \sum_{j=1}^{N} {a}_{j}\left({\gamma}_{j} - {\beta}_{j} \rho\right) + \sum_{j,k=1}^N \frac{{a}_{j,k}}{2}\left({\gamma}_{j} - {\beta}_{j} \rho\right) \left({\gamma}_{k} - {\beta}_{k} \rho\right) \right)^{2} - 4 c^{2} \prod_{j=1}^{N} \left({\gamma}_{j} - {\beta}_{j} \rho\right)^{{\beta}_{j}}
\end{equation}
and thus, provided we limit ourselves to values of $a_{j,k}$, $N$ and $\beta_j$ that ensure the right hand side of \eqref{eq:drho-sqrd-elliptic-examples} is a cubic or quartic in $\rho$, then $\rho$ is generically an elliptic function of $z$ and $u_j, v_j$ can be expressed in terms of Weierstrass $\sigma$, $\zeta$, and $\wp$. 
The system in \eqref{eq:elliptic-system} contains as special cases many well-known systems from nonlinear optics, elliptic function theory, and classical mechanics, which we summarise in the table below where, unless further specified, $Q, P$ take the most general forms given in \eqref{eq:Q-elliptic-system}.

\begin{center}
    \renewcommand{\arraystretch}{1.25}
    \begin{tabularx}{\textwidth}{|L{0.03\textwidth}|L{0.22\textwidth}|Y|}
        \hline
        \rowcolor{gray!15}
        \textbf{$N$} & \textbf{$\beta$} & \textbf{Integrable System} \\
        \hline
        4 & $\beta_j=1$ & Four-wave mixing in cubic nonlinear material \cite{heskethFWMWeierstrassElliptic2026,odzijewiczIntegrableClassicalQuantum2020,stolenParametricAmplificationFrequency1982,chenFourphotonParametricMixing1989,chenFourwaveMixingOptical1989,agrawalNonlinearFiberOptics2019}. \\
        \hline
        3 & $\beta_1=\beta_2=1$, $\beta_3=2$ & Degenerate four-wave mixing in cubic nonlinear material \cite{heskethCompleteWeierstrassElliptic2026,agrawalNonlinearFiberOptics2019}. \\
        \hline
        3 & $\beta_j=1$  & 
        \begin{tableitemize}
                \item Three-wave mixing in quadratic nonlinear material where $a_{j,k}=0$ but $a_j$ may be non-zero \cite{heskethGeneralComplexEnvelope2015,armstrongInteractionsLightWaves1962,odzijewiczIntegrabilityCorrespondenceClassical2018}.
                \item Projection from $u_j, v_j$ to the Lagrange top \cite{Whittaker1937}, where $a_{j,k}=0$ but $a_j\ne0$ and special initial conditions and parameter values apply
                \item Jacobi elliptic functions system for $\mathrm{sn}, \mathrm{cn}, \mathrm{dn}$ \cite{whittakerCourseModernAnalysis2021}, the Euler top \cite{Whittaker1937}, and a projection from $u_j, v_j$ to the Duffing oscillator \cite{kovacicJacobiEllipticFunctions2016}, where $a_j=a_{j,k}=0$ and special initial conditions apply such that $v_j \propto u_j$
        \end{tableitemize} \\
        \hline
        2 & $\beta_1=2$, $\beta_2=2$ & Orthogonal linearly polarised modes in cubic nonlinear material \cite{heskethGeneralComplexEnvelope2015,agrawalNonlinearFiberOptics2019}. \\
        \hline
        2 & $\beta_j=1$ & Asymmetric and symmetric coherent couplers in cubic nonlinear material, both Hermitian and Parity Time symmetric \cite{heskethCompleteWeierstrassElliptic2026,jensenNonlinearCoherentCoupler1982}. \\
        \hline
        2 & $\beta_1=2$, $\beta_2=1$ & Two-wave mixing in quadratic nonlinear material where $a_{j,k}=0$ but $a_j$ may be non-zero \cite{heskethGeneralComplexEnvelope2015,armstrongInteractionsLightWaves1962}. \\
        \hline
        1 & $\beta_1=4$& A system that can be thought of as the lemniscate exponential in the sense that $u_j(z)=\mathrm{cl}(z) + i\,\mathrm{sl}(z)$ and $v_j(z)=\mathrm{cl}(z) - i\,\mathrm{sl}(z)$, i.e., a version of Euler's formula for the exponential using lemniscate sine and cosine ($\mathrm{sl}, \mathrm{cl}$) \cite{whittakerCourseModernAnalysis2021}, where special initial conditions and parameter values apply. \\
        \hline
        1 & $\beta_1=3$& Dixon elliptic functions \cite{zotero-item-852} where $a_{j,k}=0$ but $a_j\ne0$ and special initial conditions and parameter values apply. \\
        \hline
    \end{tabularx}
\end{center}

\subsection{Higher-degree scalar curves and hyperelliptic solutions}
\label{sec:hyperelliptic-examples}

The elliptic restriction in the preceding examples is a restriction on the degree of the scalar polynomial, rather than on the mode count alone. Retaining $P=c\ne0$, allow arbitrary $N$, positive integer weights $\beta_j$, and a polynomial $Q$ of total degree $d$. On the conserved power line, put
\begin{equation}
    q(\rho)=Q(\gamma_1-\beta_1\rho,\ldots,\gamma_N-\beta_N\rho),
    \qquad B=\sum_{j=1}^{N}\beta_j.
\end{equation}
The same reduction gives
\begin{equation}
    \rho'^2=F(\rho),\qquad
    F(\rho)=q(\rho)^2-4c^2\prod_{j=1}^{N}(\gamma_j-\beta_j\rho)^{\beta_j}.
    \label{eq:hyperelliptic-scalar-examples}
\end{equation}
Here $\deg q\le d$, and $\deg F\le\max(2d,B)$; its actual degree must be checked after restriction and cancellation of leading terms. If $F$ is squarefree of degree $m\ge5$, the smooth compactification of $y^2=F(\rho)$ has genus $g=\lfloor(m-1)/2\rfloor\ge2$. Thus quintic and sextic scalar polynomials give genus two, while septic and octic polynomials give genus three. Repeated roots instead require normalization of a singular curve and can lower the genus.

The scalar motion is locally determined by the hyperelliptic Abelian integral
\begin{equation}
    z-z_0=\int_{\rho(z_0)}^{\rho(z)}\frac{dt}{\sqrt{F(t)}},
    \label{eq:hyperelliptic-scalar-quadrature-examples}
\end{equation}
with the sheet and continuation fixed by the initial data. Higher genus does not, in general, give a globally single-valued meromorphic inverse in $z$. In the companion manuscript \cite{zotero-item-855}, we construct local solutions for this polynomial Hamiltonian class using hyperelliptic sigma functions and Abelian integrals, for smooth odd-degree curves subject to the stated degree and regularity conditions. The scalar sigma representation lies on the one-point Abel image: only the Abel coordinate associated with $d\rho/y$ is linear in $z$, while the remaining coordinates are constrained by that image. The individual modes require further reconstruction from their powers. In that setting, \^Onishi's hyperelliptic determinant identities \cite{onishiDETERMINANTEXPRESSIONSHYPERELLIPTIC2005} encode the paired-power relations and complete weighted mode products entering the Hamiltonian, extending our genus-one connection with the Frobenius--Stickelberger determinant \cite{heskethFWMWeierstrassElliptic2026,heskethCompleteWeierstrassElliptic2026}.

\subsubsection{The \texorpdfstring{$N$}{N}-extended Euler system}
\label{sec:n-extended-euler-example}

A useful special case is the $N$-extended Euler system studied by Ferrer, Crespo, and Molero \cite{ferrerNExtendedEulerSystem2015}. Set $\beta_j=1$ and take $Q$ constant. First derive Hamilton's equations treating $u_j,v_j$ as independent canonical variables. Then choose initial data satisfying $v_j=\kappa_j u_j$, with nonzero constants obeying $\prod_j\kappa_j=-1$. Hamilton's equations preserve these proportionality relations. On this invariant family, $V=-U$, so the zero-energy condition requires $Q=0$, without requiring either product to vanish separately. The restricted equations become
\begin{equation}
    u_j'=\alpha_j\prod_{k\ne j}u_k,
    \qquad \alpha_j=-\frac{c}{\kappa_j}.
    \label{eq:n-extended-euler-reduction}
\end{equation}
Their quadratic first integrals are the restrictions of the intermodal power laws: $\kappa_j u_j^2-\kappa_k u_k^2=\gamma_j-\gamma_k$. For $N=3$, choose $\alpha_1=1$, $\alpha_2=-1$, and $\alpha_3=-k^2$. Initial data $(u_1,u_2,u_3)=(0,1,1)$ then give $(\mathrm{sn},\mathrm{cn},\mathrm{dn})$ of modulus $k$. For general $N$, the scalar polynomial in \eqref{eq:hyperelliptic-scalar-examples} reduces to $-4c^2\prod_j(\gamma_j-\rho)$, of degree $N$. With distinct $\gamma_j$, its genus is $\lfloor(N-1)/2\rfloor$: $N=4$ is still elliptic, whereas $N=5$ gives genus two. Ferrer, Crespo, and Molero examine the $N=4,5$ cases and reparameterisations that generalise the Jacobi amplitude. Their constant-$Q$, proportional-mode systems therefore sit within the broader polynomial-$Q$ canonical family considered here.

\subsubsection{A one-mode Fermat family and incomplete beta inversion}
\label{sec:fermat-beta-example}

A complementary special case has one canonical pair $u(z),v(z)$ and a single interaction weight $\beta_1=n>1$. For integers $n,m$, with $n>1$, take
\begin{equation}
    H(u,v)=\left[1-4(uv)^n\right]^m\left(u^n+v^n-1\right),
    \qquad P(x)=Q(x)=(1-4x^n)^m.
    \label{eq:fermat-beta-hamiltonian}
\end{equation}
We work on the Fermat component $u^n+v^n=1$, locally where the prefactor is finite and nonzero. A nonzero constant Fermat level can be normalized to $1$ by rescaling the modes and independent variable. For $m=0$, $P=Q=1$; the remaining integers introduce a common prefactor, polynomial for $m\ge0$ and rational for $m<0$.

The canonical scalar coordinate is $\rho(z)=-u(z)v(z)/n$. Define the fixed polynomial $D(\rho)=1-4(-n\rho)^n$. On $H=0$,
\begin{equation}
    \rho'=D(\rho)^m(u^n-v^n),\qquad
    \rho'^2=D(\rho)^{2m+1},\qquad
    u^n=\frac{1+\rho'/D(\rho)^m}{2},\quad
    v^n=\frac{1-\rho'/D(\rho)^m}{2}.
    \label{eq:fermat-beta-scalar}
\end{equation}
Compatible roots recover the modes with $uv=-n\rho$. To exhibit the inversion, choose a real local branch with $u,v>0$ and $u^n>v^n$, and put $s(z)=4[-n\rho(z)]^n$. Then $\rho=-s^{1/n}/(n4^{1/n})$ and separation gives
\begin{equation}
    B_{s(z)}\left(\frac1n,\frac12-m\right)
    -B_{s(z_0)}\left(\frac1n,\frac12-m\right)
    =-n^2 4^{1/n}(z-z_0),
    \label{eq:fermat-incomplete-beta-inversion}
\end{equation}
where $B_s(a,b)=\int_0^s t^{a-1}(1-t)^{b-1}\,dt$ is the unregularized incomplete beta function. Its local inverse determines $s(z)$, hence $\rho$ and the mode powers. For $m\ge1$, the second beta parameter is negative, but the integral remains finite for $0<s<1$.

For $m\ge1$, the scalar polynomial has repeated roots, so its high degree does not represent a generic higher-genus curve. Removing its square factor by $Y=\rho'/D(\rho)^m$ gives $Y^2=D(\rho)$, whose smooth compactification has genus $\lfloor(n-1)/2\rfloor$. The same normalization applies to the rational scalar relation for $m<0$. Thus this is a degenerate algebraic scalar family with an incomplete-beta parametrisation; its normalized scalar curve is distinct from the full Fermat curve, whose genus is $(n-1)(n-2)/2$.

\section{Log-augmented polynomial systems}
\label{sec:lotka-volterra-related}

We introduce the category of log-augmented polynomial systems to describe polynomial mode equations whose scalar generating equations retain logarithmic terms. The scalar equations belong to a broader class of rational functions augmented by powers of a logarithm, which also includes systems with rational mode equations. A useful form is
\begin{equation}
    \rho'(z)^2=R_1(\rho(z))+R_2(\rho(z))\bigl[\log R_3(\rho(z))\bigr]^m,
    \qquad m\in\mathbb Z,
    \label{eq:log-augmented-scalar}
\end{equation}
where $R_1,R_2,R_3$ are fixed rational functions of their scalar argument and $m$ is a constant integer. More generally, several logarithmic powers may occur together. We work locally on consistent logarithmic branches, away from poles and other singularities. Such equations can arise in the Hamiltonian framework considered here when logarithmic terms $[\log \mathcal R(x_1,\ldots,x_N)]^M$ are combined with rational functions of the modal powers $x_j=u_jv_j$ and the weighted products $U=\prod_j u_j^{\beta_j}$, $V=\prod_j v_j^{\beta_j}$, for choices compatible with the scalar reduction. The Hamiltonian exponent $M$ need not equal $m$: squaring or eliminating the mode products can generate mixed logarithmic powers.

These systems are interesting because their equations of motion can be polynomial or rational even though their scalar reductions contain logarithms. Differentiation of logarithmic Hamiltonian terms produces rational expressions, while the zero-energy relation can eliminate remaining logarithms from the mode equations. The logarithmic dependence nevertheless persists in the conserved relation and reappears in the scalar equation for $\rho$. Thus, unlike the systems in Section~\ref{sec:elliptic-examples}, polynomial mode equations here need not imply an elliptic or hyperelliptic scalar reduction: the logarithmic terms generally prevent the reduced relation from defining an algebraic curve. The examples below illustrate how this structure can remain hidden in otherwise familiar-looking polynomial systems.

We first present an $m=1$ reduction of a polynomial system considered by Sergyeyev \cite{sergyeyevCouplingConstantMetamorphosis2012} and a separate rational-coefficient equation with an explicit Lambert $W$ solution. We then consider four systems involving squared logarithmic expressions, possibly with shifted or rescaled arguments and lower logarithmic powers after expansion.

\subsection{First-power logarithmic terms: \texorpdfstring{$m=1$}{m=1}}
\label{sec:log-linear-examples}

\subsubsection{A polynomial system considered by Sergyeyev}
\label{sec:sergyeyev-log-example}

In Example 2 of \cite{sergyeyevCouplingConstantMetamorphosis2012}, Sergyeyev applies coupling constant metamorphosis to a system attributed to Goriely and obtains the polynomial system for $X(z),Y(z),Z(z)$, with constant parameter $b$:
\begin{align}
    X'&=-2X^2+2Z, & Y'&=-3XY, & Z'&=-2XZ+2X^3+6bX.
    \label{eq:sergyeyev-polynomial}
\end{align}
Its first integrals have constant values $I,J$:
\begin{equation}
    I=\frac{b+X^2-Z}{Y^2},\qquad
    J=\frac{X^2}{2}+Z+b\log(Y^2).
    \label{eq:sergyeyev-integrals}
\end{equation}
Sergyeyev uses these to reduce the system to a scalar equation containing a first power of a logarithm. To relate it to the present mode variables, choose a local square root $Z(z)=\widehat Z(z)^2$ and set
\begin{equation}
    u(z)=X(z)-\widehat Z(z),\qquad v(z)=X(z)+\widehat Z(z),\qquad
    x(z)=u(z)v(z)=X(z)^2-Z(z).
\end{equation}
For $I\ne0$, the integral $J$ becomes
\begin{equation}
    J=\frac38(u^2+v^2)-\frac{uv}{4}
      +b\log\left(\frac{uv+b}{I}\right).
\end{equation}
The first integral defines a fixed function $g$ of the modal power, whose value $g(u(z)v(z))$ varies along the trajectory:
\begin{equation}
    g(uv)=\sqrt{\frac83J-\frac43uv-\frac{8b}{3}
       \log\left(\frac{uv+b}{I}\right)}.
    \label{eq:sergyeyev-power-prefactor}
\end{equation}
Choose its local branch so that $g(uv)=v-u$ on the conserved surface, away from its zeros and the logarithmic singularities. A single Hamiltonian of the form \eqref{eq:Hamiltonian}, with $N=1$ and $\beta_1=2$, is then
\begin{equation}
    H(u,v)=\frac{uv+b}{g(uv)}\left[
       -\frac32(u^2+v^2)+uv+4J
       -4b\log\left(\frac{uv+b}{I}\right)
    \right].
    \label{eq:sergyeyev-power-hamiltonian}
\end{equation}
On $H=0$, its canonical equations reproduce the transformed $(X,Z)$ subsystem in the original independent variable $z$. The remaining variable is recovered from $Y^2=(uv+b)/I$ with a consistent branch. We give this Hamiltonian representation for completeness of the examples; we make no particular claim that it offers an advantage over Sergyeyev's original formulation (beyond its symmetry in $u,v$).

In the original variable $z$, direct differentiation gives $(uv)'=-3(uv+b)(u+v)$. Our one-mode convention with $\beta_1=2$ fixes $\gamma_1=0$ and hence
\begin{equation}
    \rho(z)=-\frac{u(z)v(z)}{2}.
\end{equation}
Using the first integral to eliminate $(u+v)^2$ then gives Sergyeyev's equation
\begin{equation}
    \rho'^2=6(b-2\rho)^2\left[
        J-2\rho-b\log\left(\frac{b-2\rho}{I}\right)
    \right].
    \label{eq:sergyeyev-log-scalar}
\end{equation}
This is an $m=1$ member of \eqref{eq:log-augmented-scalar}, with polynomial $R_1,R_2$ and $R_3(\rho)=(b-2\rho)/I$, retaining the canonical modal-power coordinate $\rho$ and the original independent variable $z$.

\subsubsection{An explicit Lambert \texorpdfstring{$W$}{W} example}
\label{sec:lambert-log-example}

Independently of the preceding system, let $a$ be a nonzero constant and consider the rational-coefficient equation for $\rho(z)$
\begin{equation}
    \rho'^2=\frac{4\rho^2}{(a\rho-1)^2}\left(a\rho-\log \rho\right),
    \qquad a\ne0.
    \label{eq:lambert-log-rational-ode}
\end{equation}
A local solution family is
\begin{equation}
    \rho(z)=-\frac1a\,
    W_k\left(-a e^{-(z-z_0)^2}\right),
    \label{eq:lambert-log-explicit-solution}
\end{equation}
where $z_0$ is a constant of integration, $k$ is a fixed branch index, and $W_k$ is the corresponding branch of the Lambert function \cite{corlessLambertWFunction1996}. Indeed, its defining identity yields
\begin{equation}
    a\rho-\log \rho=(z-z_0)^2,\qquad
    \rho'=\frac{2(z-z_0)\rho}{a\rho-1},
\end{equation}
which verify \eqref{eq:lambert-log-rational-ode}. The logarithmic branch is chosen to satisfy the first identity, and the formula is used locally away from $a\rho=1$ and other branch singularities.

This scalar equation also supplies an explicit one-mode Hamiltonian example. With $\rho=-uv$, consider
\begin{equation}
    H(u,v)=\frac{\sqrt a\,uv}{1+a uv}
       \left[u+v-2\sqrt{-\frac{\log(-uv)}a}\right].
    \label{eq:lambert-log-mode-hamiltonian}
\end{equation}
Differentiating before restricting to $H=0$, and then using the zero-energy relation to eliminate the logarithm, gives
\begin{align}
    u'&=\frac{u[2+a v(u+v)]}{\sqrt a(1+a uv)(u+v)},\notag\\
    v'&=-\frac{v[2+a u(u+v)]}{\sqrt a(1+a uv)(u+v)}.
    \label{eq:lambert-log-mode-equations}
\end{align}
The mode solutions are completely explicit:
\begin{equation}
\begin{aligned}
    u(z)&=\frac{z-z_0}{\sqrt a}
        +\sqrt{\frac{(z-z_0)^2+W_k\!\left(-a e^{-(z-z_0)^2}\right)}{a}},\\
    v(z)&=-\frac{z-z_0}{\sqrt a}
        +\sqrt{\frac{(z-z_0)^2+W_k\!\left(-a e^{-(z-z_0)^2}\right)}{a}}.
\end{aligned}
    \label{eq:lambert-log-explicit-modes}
\end{equation}
Indeed, the rational system gives $(u-v)'=2/\sqrt a$ and conserves
\begin{equation}
    C=\frac a4(u+v)^2+\log(-uv).
    \label{eq:lambert-log-mode-integral}
\end{equation}
Its zero level yields $a\rho-\log\rho=(z-z_0)^2$, recovering the stated Lambert-$W$ solution. Roots and logarithms are chosen consistently with the selected $H=0$ branch, away from $uv(1+a uv)(u+v)=0$.

\subsection{Squared logarithmic terms: \texorpdfstring{$m=2$}{m=2}}
\label{sec:log-quadratic-examples}

The elliptic examples in Section~\ref{sec:elliptic-examples} are generated by scalar differential equations which can be reduced to the Weierstrass equation for $s(z)$, with constant invariants $g_2,g_3$
\begin{equation}
    s'^2=4s^3-g_2s-g_3.
\end{equation}
We now consider systems generated by a logarithmic deformation of this equation. A convenient factorised form of the deformation for $\rho(z)$, with constant coefficients $c_0,c_1,c_2,c_3,c_5$, is
\begin{equation}
    \rho'^2
    =
    c_0(\rho-c_1)(\rho-c_2)(\rho-c_3)
    +
    c_5\rho^2\log^2 \rho. \label{eq:log-weierstrass-factorised}
\end{equation}
Under the linear change of variable
\begin{equation}
    \rho(z)=\frac{4s(z)}{c_0}+\frac{c_1+c_2+c_3}{3}
    =\frac{4(s(z)+c_6)}{c_0},
\end{equation}
with $c_0\ne0$ and the constant $c_6=(c_0/12)(c_1+c_2+c_3)$, equation \eqref{eq:log-weierstrass-factorised} takes the Weierstrass-like form
\begin{equation}
    s'^2
    =
    4s^3-g_2s-g_3
    +
    c_5(s+c_6)^2
    \log^2\left(\frac{4(s+c_6)}{c_0}\right). \label{eq:log-weierstrass-normal}
\end{equation}
We refer to \eqref{eq:log-weierstrass-normal} as the logarithmically deformed Weierstrass equation. Setting $c_5=0$ recovers the ordinary Weierstrass equation. It is also useful to note that the reciprocal change $\xi(z)=1/\rho(z)$ gives
\begin{equation}
    \xi'^2
    =
    -c_0\xi\prod_{j=1}^{3}(c_j\xi-1)
    +
    c_5\xi^2\log^2\xi. \label{eq:log-weierstrass-reciprocal}
\end{equation}
We will show that the Lotka--Volterra model, a parity-time symmetric dimer, and two further mode systems are related through special or degenerate cases of these equations. These equations also admit a local representation through a Weierstrass function after reparameterisation, as detailed in the supplementary computational notebooks. The reparameterisation requires inversion of a generally nonelliptic integral, so this representation does not make the solutions elliptic functions of the original independent variable.

\subsubsection{The Lotka--Volterra model}

Let $\widehat u(z),\widehat v(z)$ satisfy the two-species Lotka--Volterra equations. The parameters $\alpha,\beta,\kappa,\delta$ are constant, and we use the sign convention
\begin{align}
    \widehat{u}' &= \alpha\widehat{u}-\delta\widehat{u}\widehat{v}, \notag \\
    \widehat{v}' &= -\beta\widehat{v}-\kappa\widehat{u}\widehat{v}.
\end{align}
For $\kappa,\delta\ne0$, introduce the rescaled variables
\begin{equation}
    \widehat{u}(z)=\frac{u(z)}{\kappa},
    \qquad
    \widehat{v}(z)=-\frac{v(z)}{\delta},
\end{equation}
which satisfy
\begin{align}
    u' &= \alpha u+uv, \notag \\
    v' &= -\beta v-uv. \label{eq:lotka-volterra-rescaled}
\end{align}
The rescaled system conserves
\begin{equation}
    \mathcal{K}(u,v)
    =
    \alpha\log(-v)+\beta\log u+u+v.
\end{equation}
Fixing a value $K$ of this conserved quantity, define the canonical Hamiltonian
\begin{equation}
    H_{\mathrm{LV}}
    =
    uv\left[
        \alpha\log(-v)+\beta\log u+u+v-K
    \right]. \label{eq:lotka-volterra-hamiltonian}
\end{equation}
Hamilton's equations give
\begin{align}
    u' &= \partial_vH_{\mathrm{LV}}
    =\frac{H_{\mathrm{LV}}}{v}+\alpha u+uv, \notag \\
    v' &= -\partial_uH_{\mathrm{LV}}
    =-\frac{H_{\mathrm{LV}}}{u}-\beta v-uv.
\end{align}
Thus, locally where $uv\ne0$, the zero-energy level $H_{\mathrm{LV}}=0$ reproduces \eqref{eq:lotka-volterra-rescaled}.

In the symmetric case $\beta=\alpha$, the Hamiltonian becomes
\begin{equation}
    H_{\mathrm{LV}}
    =
    (u+v)uv
    -
    \left(K-\alpha\log(-uv)\right)uv.
\end{equation}
This is the $N=1$, $\beta_1=1$ case of \eqref{eq:Hamiltonian}. Writing $x(z)=u(z)v(z)$, the fixed power-dependent functions are
\begin{equation}
    P(x)=x,
    \qquad
    Q(x)=\left(K-\alpha\log(-x)\right)x.
\end{equation}
Our convention $\sum_{j=1}^{N}\gamma_j=0$ fixes the additive freedom in $\rho$. For $N=1$ this necessarily gives $\gamma_1=0$, so that $\rho=-uv$, and equation \eqref{eq:drho-sqrd-ham} becomes
\begin{equation}
    \rho'^2
    =
    4\rho^3
    +
    \rho^2\left(K-\alpha\log\rho\right)^2. \label{eq:rho-lotka-volterra}
\end{equation}
For $\alpha\ne0$, this is a special case of \eqref{eq:log-weierstrass-normal}: one may take $g_2=g_3=c_6=0$, $c_5=\alpha^2$, and choose the scale inside the logarithm so that $c_0=4e^{K/\alpha}$.

\subsubsection{A parity-time symmetric dimer}

Ramezani et al. \cite{ramezaniUnidirectionalNonlinearPTsymmetric2010} study a nonlinear optical dimer with balanced gain and loss and cubic Kerr nonlinearity. Its polynomial mode equations conceal the inverse-trigonometric conservation law that produces the logarithmic scalar reduction below. We treat $\psi_1(z),\psi_2(z),\widetilde\psi_1(z),\widetilde\psi_2(z)$ as four independent complex variables: the tilde labels a separate variable and does not denote complex conjugation. With constant parameters $\chi,\kappa$, the complexified equations are:
\begin{align}
    \psi_1' &= \mathrm{i}\chi\psi_1^2\widetilde{\psi}_1+\kappa\psi_1+\mathrm{i}\psi_2, \notag \\
    \psi_2' &= \mathrm{i}\chi\psi_2^2\widetilde{\psi}_2-\kappa\psi_2+\mathrm{i}\psi_1, \notag \\
    \widetilde{\psi}_1' &= -\mathrm{i}\chi\psi_1\widetilde{\psi}_1^2+\kappa\widetilde{\psi}_1-\mathrm{i}\widetilde{\psi}_2, \notag \\
    \widetilde{\psi}_2' &= -\mathrm{i}\chi\psi_2\widetilde{\psi}_2^2-\kappa\widetilde{\psi}_2-\mathrm{i}\widetilde{\psi}_1. \label{eq:ramezani-system}
\end{align}
The physical dimer is recovered by $\widetilde\psi_j=\psi_j^*$; $\chi$ is the Kerr coefficient and $\kappa$ the balanced gain/loss coefficient, with the linear coupling normalized to unity.
Introduce the complex Stokes variables $S_0(z),S_1(z),S_2(z)\in\mathbb C$, which become real on the physical conjugate restriction:
\begin{align}
    S_0
    &=
    \psi_1\widetilde{\psi}_1
    +
    \psi_2\widetilde{\psi}_2,
    \notag \\
    S_1
    &=
    \psi_1\widetilde{\psi}_2
    +
    \psi_2\widetilde{\psi}_1,
    \notag \\
    S_2
    &=
    \mathrm{i}
    \left(
        \psi_1\widetilde{\psi}_2
        -
        \psi_2\widetilde{\psi}_1
    \right).
\end{align}
In the notation used here, the first integrals of \cite{ramezaniUnidirectionalNonlinearPTsymmetric2010}, together with a locally equivalent arctangent representation, are
\begin{align}
    C^2
    &=
    (\chi S_1-2)^2+(\chi S_2)^2,
    \notag \\
    J
    &=
    S_0
    +
    \frac{2\kappa}{\chi}
    \arcsin\left(\frac{\chi S_1-2}{C}\right)
    \notag \\
    &=
    S_0
    -
    \frac{2\kappa}{\chi}
    \arctan\left(
        \frac{\chi S_1-2}{\chi S_2}
    \right).
    \label{eq:ramezani-original-integrals}
\end{align}
For real variables, the principal inverse-sine expression generally requires $C\chi S_2<0$ for conservation; extension beyond this sign domain requires continuation of its branch. The cited paper specifies initial data and signed values of $C$, but does not state this branch restriction explicitly. The arctangent form likewise requires a continued angle. For the complexified system, a logarithmic representation makes the conservation law more direct:
\begin{align}
    C^2
    &=
    4
    \left(
        \chi\psi_1\widetilde{\psi}_2-1
    \right)
    \left(
        \chi\psi_2\widetilde{\psi}_1-1
    \right),
    \notag \\
    J
    &=
    S_0
    -
    \frac{2\mathrm{i}\kappa}{\chi}
    \log\left(
        \frac{\mathrm{i}C}
        {2(\chi\psi_1\widetilde{\psi}_2-1)}
    \right).
    \label{eq:ramezani-log-integrals}
\end{align}
With compatible branches, \eqref{eq:ramezani-log-integrals} recovers the local inverse-trigonometric forms in \eqref{eq:ramezani-original-integrals}. Direct differentiation gives $J\,'=0$ for the complexified equations wherever the logarithmic expression is defined, without the real sign restriction above; the logarithm must be continued along the trajectory rather than reset to its principal branch. In its logarithmic form, $J$ has the same structure as the Lotka--Volterra conservation law: a sum of modal powers together with a logarithm.

We now relate \eqref{eq:ramezani-system} to a canonical Hamiltonian in the coordinates of the present theory. For one choice of the square-root branch, introduce canonical modes $u_j(z),v_j(z)$, $j=1,2$, and consider the $N=2$, $\beta_1=\beta_2=1$ Hamiltonian
\begin{multline}
    H_{\mathrm{PT}}
    =
    \frac{\sqrt{2}}{4}
    (u_1u_2+v_1v_2)
    \sqrt{C^2-4u_1v_1-4u_2v_2+4}
    \\
    +
    \frac{\mathrm{i}}{8}
    (C^2-4u_1v_1-4u_2v_2+4)
    \left[
        J\chi+2\mathrm{i}\kappa
        \log\left(
            \frac{-4\mathrm{i}C}
            {C^2-4u_1v_1-4u_2v_2+4}
        \right)
    \right]. \label{eq:ramezani-canonical-hamiltonian}
\end{multline}
The associated intermodal power law may be written
\begin{equation}
    C^2=4u_1v_1-4u_2v_2+4.
\end{equation}
On $H_{\mathrm{PT}}=0$, define a phase function $\varphi(z)$ by
\begin{equation}
    \varphi'
    =
    -\frac{3\mathrm{i}}{4}
    \left[
        J\chi+2\mathrm{i}\kappa
        \log\left(
            \frac{-4\mathrm{i}C}
            {C^2-4u_1v_1-4u_2v_2+4}
        \right)
    \right],
\end{equation}
and make the coordinate transformation
\begin{align}
    \psi_1
    &=
    \frac{v_2e^{-\varphi}}{\sqrt{\chi}},
    &
    \psi_2
    &=
    \frac{2\mathrm{i}u_1e^{-\varphi}}
    {\sqrt{\chi}\sqrt{C^2-4u_1v_1}},
    \notag \\
    \widetilde{\psi}_1
    &=
    \frac{2\mathrm{i}v_1e^{\varphi}}
    {\sqrt{\chi}\sqrt{C^2-4u_1v_1}},
    &
    \widetilde{\psi}_2
    &=
    \frac{u_2e^{\varphi}}{\sqrt{\chi}}. \label{eq:ramezani-coordinate-transform}
\end{align}
Direct substitution of Hamilton's equations for \eqref{eq:ramezani-canonical-hamiltonian} into \eqref{eq:ramezani-coordinate-transform} gives \eqref{eq:ramezani-system}.

As usual, set $x_j=u_jv_j$ and
\begin{equation}
    \rho=-\frac{x_1+x_2}{2},
    \qquad
    x_j=\gamma_j-\rho,
\end{equation}
where the intermodal power law fixes
\begin{equation}
    \gamma_1=\frac{C^2}{8}-\frac{1}{2},
    \qquad
    \gamma_2=-\gamma_1.
\end{equation}
The scalar equation generated by \eqref{eq:ramezani-canonical-hamiltonian} is
\begin{multline}
    \rho'^2
    =
    -\frac{(C^2+8\rho+4)^2}{64}
    \left[
        J\chi+2\mathrm{i}\kappa
        \log\left(\frac{-4\mathrm{i}C}{C^2+8\rho+4}\right)
    \right]^2
    \\
    +
    \frac{(C^2-8\rho-4)(C^2+8\rho-4)(C^2+8\rho+4)}{128}.
    \label{eq:rho-ramezani}
\end{multline}
Define the function $w(z)$ and the constant $A$ by
\begin{equation}
    w(z)=C^2+8\rho(z)+4,
    \qquad
    A=-4\mathrm{i}C\exp\left(\frac{J\chi}{2\mathrm{i}\kappa}\right),
\end{equation}
this becomes
\begin{equation}
    w'^2
    =
    -\frac{1}{2}w(w-8)(w-2C^2)
    +
    4\kappa^2w^2\log^2\left(\frac{w}{A}\right), \label{eq:ramezani-log-weierstrass}
\end{equation}
which is again of the form \eqref{eq:log-weierstrass-factorised}, up to a constant rescaling of its argument.

The connection between \eqref{eq:rho-lotka-volterra}, \eqref{eq:rho-ramezani}, and the logarithmically deformed Weierstrass equation suggests further mode systems of the same type.

\subsubsection{A three-mode logarithmic system}

Take $N=3$, $\beta_j=1$, and modes $u_j(z),v_j(z)$ with modal powers $x_j(z)=u_j(z)v_j(z)$. Let $a_0,a_1,a_2,a_3$ be constants, and define
\begin{equation}
    \rho
    =
    -\frac{u_1v_1+u_2v_2+u_3v_3}{3}.
\end{equation}
Define the canonical Hamiltonian
\begin{multline}
    H_3
    =
    \left[
        -\frac{a_0}{3}
        (u_1v_1+u_2v_2+u_3v_3)
        -a_1
    \right]
    \\
    {}\times
    \log\left[
        a_3
        -
        \frac{a_2}{3}
        (u_1v_1+u_2v_2+u_3v_3)
    \right]
    +
    u_1u_2u_3+v_1v_2v_3. \label{eq:three-mode-log-hamiltonian}
\end{multline}
With constants $\gamma_j$ fixed by the initial powers, this has the two independent intermodal power laws
\begin{align}
    u_1v_1-u_2v_2
    &=
    \gamma_1-\gamma_2,
    \notag \\
    u_1v_1-u_3v_3
    &=
    \gamma_1-\gamma_3,
\end{align}
and on $H_3=0$ generates
\begin{equation}
    \rho'^2
    =
    (a_0\rho-a_1)^2\log^2(a_2\rho+a_3)
    -
    4(\gamma_1-\rho)(\gamma_2-\rho)(\gamma_3-\rho).
    \label{eq:three-mode-log-rho}
\end{equation}
Equation \eqref{eq:three-mode-log-rho} is an $m=2$ member of the log-augmented class \eqref{eq:log-augmented-scalar} without any parameter restriction. The choice $a_1=a_3=0$ puts it directly in the factorised form \eqref{eq:log-weierstrass-factorised}. More generally, that form is reached by the affine change $w=a_2\rho+a_3$ whenever $a_0a_2\ne0$ and $a_0a_3+a_1a_2=0$, so that the two linear factors have the same zero.

For the choice $a_1=a_3=0$, introduce new functions $\widetilde{u}_j(z),\widetilde{v}_j(z)$ and $S(z)$, locally where $S(z)\ne0$, by
\begin{equation}
    u_j(z)=S(z)\widetilde{u}_j(z),
    \qquad
    v_j(z)=S(z)\widetilde{v}_j(z),
    \qquad
    S
    =
    \widetilde{u}_1\widetilde{v}_1
    +
    \widetilde{u}_2\widetilde{v}_2
    +
    \widetilde{u}_3\widetilde{v}_3.
\end{equation}
Then $\rho=-S^3/3$, and the transformed equations take the polynomial form
\begin{align}
    \widetilde{u}_1'
    &=
    \left(
        -\widetilde{u}_1\widetilde{v}_1
        +\widetilde{u}_2\widetilde{v}_2
        +\widetilde{u}_3\widetilde{v}_3
    \right)
    \widetilde{v}_2\widetilde{v}_3
    -
    \frac{a_0}{3}\widetilde{u}_1,
    \notag \\
    \widetilde{u}_2'
    &=
    \left(
        \widetilde{u}_1\widetilde{v}_1
        -\widetilde{u}_2\widetilde{v}_2
        +\widetilde{u}_3\widetilde{v}_3
    \right)
    \widetilde{v}_1\widetilde{v}_3
    -
    \frac{a_0}{3}\widetilde{u}_2,
    \notag \\
    \widetilde{u}_3'
    &=
    \left(
        \widetilde{u}_1\widetilde{v}_1
        +\widetilde{u}_2\widetilde{v}_2
        -\widetilde{u}_3\widetilde{v}_3
    \right)
    \widetilde{v}_1\widetilde{v}_2
    -
    \frac{a_0}{3}\widetilde{u}_3,
    \notag \\
    \widetilde{v}_1'
    &=
    -\left(
        -\widetilde{u}_1\widetilde{v}_1
        +\widetilde{u}_2\widetilde{v}_2
        +\widetilde{u}_3\widetilde{v}_3
    \right)
    \widetilde{u}_2\widetilde{u}_3
    +
    \frac{a_0}{3}\widetilde{v}_1,
    \notag \\
    \widetilde{v}_2'
    &=
    -\left(
        \widetilde{u}_1\widetilde{v}_1
        -\widetilde{u}_2\widetilde{v}_2
        +\widetilde{u}_3\widetilde{v}_3
    \right)
    \widetilde{u}_1\widetilde{u}_3
    +
    \frac{a_0}{3}\widetilde{v}_2,
    \notag \\
    \widetilde{v}_3'
    &=
    -\left(
        \widetilde{u}_1\widetilde{v}_1
        +\widetilde{u}_2\widetilde{v}_2
        -\widetilde{u}_3\widetilde{v}_3
    \right)
    \widetilde{u}_1\widetilde{u}_2
    +
    \frac{a_0}{3}\widetilde{v}_3.
    \label{eq:three-mode-log-polynomial}
\end{align}
Each nonlinear term is a product of four mode amplitudes, making the resemblance to a five-wave-mixing system visible directly from the equations.
The intermodal power laws become
\begin{align}
    \left(
        \widetilde{u}_1\widetilde{v}_1
        -
        \widetilde{u}_2\widetilde{v}_2
    \right)S^2
    =
    \gamma_1-\gamma_2,
    \notag \\
    \left(
        \widetilde{u}_1\widetilde{v}_1
        -
        \widetilde{u}_3\widetilde{v}_3
    \right)S^2
    =
    \gamma_1-\gamma_3.
\end{align}

\subsubsection{A two-mode logarithmic system}

For a second example, take $N=2$, $\beta_1=\beta_2=1$, modes $u_1(z),u_2(z),v_1(z),v_2(z)$, and constant parameters $a_0,a_1,a_2$. Define the canonical Hamiltonian
\begin{equation}
    H_2
    =
    (a_1u_1v_1+a_2u_2v_2)
    \left[
        \log\left(
            a_0(a_1u_1v_1+a_2u_2v_2)
        \right)
        -
        u_1u_2-v_1v_2
    \right]. \label{eq:two-mode-log-hamiltonian}
\end{equation}
On $H_2=0$, Hamilton's equations simplify to the polynomial system
\begin{align}
    u_1'
    &=
    a_1u_1
    -
    (a_1u_1v_1+a_2u_2v_2)v_2,
    \notag \\
    u_2'
    &=
    a_2u_2
    -
    (a_1u_1v_1+a_2u_2v_2)v_1,
    \notag \\
    v_1'
    &=
    (a_1u_1v_1+a_2u_2v_2)u_2
    -
    a_1v_1,
    \notag \\
    v_2'
    &=
    (a_1u_1v_1+a_2u_2v_2)u_1
    -
    a_2v_2.
    \label{eq:two-mode-log-polynomial}
\end{align}
The nonlinear terms are cubic in the mode amplitudes, as in four-wave mixing in a cubic medium.
Writing $x_1=u_1v_1$ and $x_2=u_2v_2$, the modal powers satisfy $x_1'-x_2'=0$. Taking
\begin{equation}
    \rho=-\frac{x_1+x_2}{2},
\end{equation}
we may therefore write
\begin{equation}
    x_j=\gamma_j-\rho,
    \qquad
    \gamma_1+\gamma_2=0.
\end{equation}
For compactness in the scalar equation only, define the fixed affine function $\ell$ by
\begin{equation}
    \ell(\rho)
    =
    a_1(\gamma_1-\rho)
    +
    a_2(\gamma_2-\rho).
\end{equation}
Equation \eqref{eq:drho-sqrd-ham} gives
\begin{equation}
    \rho'^2
    =
    \ell(\rho)^2
    \left[
        \log^2(a_0\ell(\rho))
        -
        4(\gamma_1-\rho)(\gamma_2-\rho)
    \right]. \label{eq:two-mode-log-rho}
\end{equation}
Finally, on defining
\begin{equation}
    w(z)=\frac{1}{a_0\ell(\rho(z))},
\end{equation}
equation \eqref{eq:two-mode-log-rho} becomes
\begin{equation}
    w'^2
    =
    \frac{4}{a_0^2}
    \left(2a_0a_1\gamma_1w-1\right)
    \left(2a_0a_2\gamma_1w+1\right)
    +
    (a_1+a_2)^2w^2\log^2w. \label{eq:two-mode-log-weierstrass}
\end{equation}
This is a degenerate case of the factorised form \eqref{eq:log-weierstrass-factorised}, in which one root is taken to infinity with a compensating rescaling of the cubic coefficient.

These examples show how logarithmic scalar generating equations connect known systems and generate further mode equations. The scalar equation, the canonical coordinates, and the choice of independent variable must all be specified when comparing their solutions.

\section{Rational generating equations and generalised Lambert--W-type inverse laws}
\label{sec:rational-generating-equations}

Rational choices of the generating equation form a simple class for which the inverse-function viewpoint can be made explicit.
We first record the correspondence between these equations and a broad family of generally multivalued inverse laws.

\begin{theorem}[Rational generating equations]
Let
\begin{equation}
    \rho'=f(\rho)=\frac{A(\rho)}{B(\rho)},
    \label{eq:rational-generating-ode}
\end{equation}
where $A$ and $B$ are coprime polynomials and $f$ is not identically zero.
Every nonconstant solution is locally the inverse of a relation of the form
\begin{equation}
    e^{\mathcal{P}(\rho)+\mathcal{R}(\rho)}
    \prod_{\ell=1}^{L}(\rho-a_\ell)^{\nu_\ell}
    =e^z,
    \label{eq:extended-lambert-inverse-law}
\end{equation}
where $\mathcal{P}$ is a polynomial whose constant term is fixed by the initial condition, and $\mathcal{R}$ is a rational function consisting of a finite sum of principal parts.
Conversely, every local inverse defined by a relation of the form \eqref{eq:extended-lambert-inverse-law} satisfies an autonomous differential equation in which $f(\rho)$ is rational.
\end{theorem}

\begin{proof}
On a region where $f(\rho)$ is finite and nonzero, interchange the dependent and independent variables to obtain
\begin{equation}
    \frac{dz}{d\rho}=\frac{B(\rho)}{A(\rho)}.
\end{equation}
Over $\mathbb{C}$, its partial-fraction expansion may be written
\begin{equation}
    \frac{B(\rho)}{A(\rho)}
    =p(\rho)
    +\sum_{\ell=1}^{L}\sum_{k=1}^{n_\ell}
    \frac{c_{\ell k}}{(\rho-a_\ell)^k},
    \label{eq:rational-ode-partial-fractions}
\end{equation}
where $p$ is a polynomial and the positive integer $n_\ell$ is the order of the pole at $a_\ell$.
Termwise integration gives
\begin{equation}
    z
    =\mathcal{P}(\rho)
    +\sum_{\ell=1}^{L}c_{\ell1}\log(\rho-a_\ell)
    -\sum_{\ell=1}^{L}\sum_{k=2}^{n_\ell}
    \frac{c_{\ell k}}{(k-1)(\rho-a_\ell)^{k-1}},
    \label{eq:rational-ode-integrated}
\end{equation}
where $\mathcal{P}'=p$ and the constant of integration has been included in the constant term of $\mathcal{P}$.
Exponentiating \eqref{eq:rational-ode-integrated} gives \eqref{eq:extended-lambert-inverse-law}, with
\begin{equation}
    \nu_\ell=c_{\ell1},
    \qquad
    \mathcal{R}(\rho)
    =-\sum_{\ell=1}^{L}\sum_{k=2}^{n_\ell}
    \frac{c_{\ell k}}{(k-1)(\rho-a_\ell)^{k-1}}.
\end{equation}
For the converse, take a local logarithm of \eqref{eq:extended-lambert-inverse-law} and differentiate to find
\begin{equation}
    \rho'
    =\left[
        \mathcal{P}'(\rho)
        +\mathcal{R}'(\rho)
        +\sum_{\ell=1}^{L}\frac{\nu_\ell}{\rho-a_\ell}
    \right]^{-1},
    \label{eq:extended-lambert-rational-ode}
\end{equation}
which is rational in $\rho$.
\end{proof}

We refer to the generally multivalued local inverses in \eqref{eq:extended-lambert-inverse-law} as extended Lambert--W-type inverse functions.
The classical Lambert function is recovered by taking $\mathcal{P}(\rho)=\rho$, $\mathcal{R}=0$, $L=1$, $a_1=0$, and $\nu_1=1$, so that \eqref{eq:extended-lambert-inverse-law} reduces to $\rho e^\rho=e^z$ \cite{corlessLambertWFunction1996}; an arbitrary multiplicative factor on the right is equivalent to adding a constant to $\mathcal{P}$.
The generalised Lambert function of Mez\H{o} and Baricz is also contained in \eqref{eq:extended-lambert-inverse-law}: it corresponds to a linear $\mathcal{P}$, $\mathcal{R}=0$, and integer powers of the linear factors \cite{mezoGeneralizationLambert$W$2015}.
Castle subsequently allowed arbitrary complex powers of the linear factors and developed Taylor series for the resulting inverse functions \cite{castleTaylorSeriesGeneralized2018}.
For this subfamily, a Taylor series $w(Z)=\sum_{n=0}^{\infty}b_nZ^n$ for the inverse function gives the corresponding solution of the rational equation as
\begin{equation}
    \rho(z)=w(e^z)
    =\sum_{n=0}^{\infty}b_ne^{nz},
    \label{eq:castle-series-rational-ode}
\end{equation}
where the series is used within its domain of convergence.
The further inclusion of a general polynomial $\mathcal{P}$ and rational principal parts $\mathcal{R}$ in the exponential is what permits the partial-fraction expansion of an arbitrary rational equation to be represented by \eqref{eq:extended-lambert-inverse-law}.
This family is not intended to include every inverse function that has been described as a generalisation of Lambert $W$; Subsection \ref{sec:inverse-spherical-bessel-example} considers a different extension based on inverse spherical Bessel functions.
As elsewhere in this paper, the logarithms, powers, and inverse functions are understood locally with branches fixed by the initial condition.
Zeros of $f$ also give constant solutions of \eqref{eq:rational-generating-ode}; these are separate from the nonconstant locally invertible solutions described by the theorem.

A first consequence is that rational functions of $\rho(z)$ have elementary antiderivatives along a solution of \eqref{eq:rational-generating-ode}.
Indeed, if $G$ is rational, then
\begin{equation}
    \int G(\rho(z))\,dz
    =\int \frac{G(\rho)}{f(\rho)}\,d\rho,
    \label{eq:rational-rho-quadrature}
\end{equation}
and the integrand on the right is rational.
Over $\mathbb{C}$ the result is therefore a sum of a rational function of $\rho$ and logarithms of linear functions of $\rho$.
This is particularly useful when solving the integrable systems generated in this way: once $\rho(z)$ is known, any rational functions of $\rho$ that occur in the remaining separated mode equations can be integrated explicitly in elementary form.

There is also a direct consequence for the coupled systems constructed above: rational choices of $F$ and $P$ generate coupled first-order equations that are rational in the modes.
Suppose that $F(x_1,\dots,x_N)$ in \eqref{eq:f-F} and the freely chosen function $P(x_1,\dots,x_N)$ are rational.
Since the $\beta_j$ are integers, $R=\prod_m x_m^{\beta_m}$ and the modal products $\prod_m u_m^{\beta_m}$ and $\prod_m v_m^{\beta_m}$ are rational functions of the modes.
Although $Q$ in \eqref{eq:Q-F-P-R} need not itself be rational away from the chosen energy level, on $H=0$ we have
\begin{equation}
    Q
    =P\left(
        \prod_{m=1}^{N}u_m^{\beta_m}
        +\prod_{m=1}^{N}v_m^{\beta_m}
    \right).
    \label{eq:rational-system-Q-on-H-zero}
\end{equation}
Using \eqref{eq:rational-system-Q-on-H-zero} to remove the square root from the equations obtained through \eqref{eq:Q-F-P-R}, we obtain on any regular part of $H=0$ a system of rational ordinary differential equations for $u_j,v_j$.
This implication is only in one direction: a coupled system that is rational or polynomial in the modes need not have a rational generating function $f(\rho)$ and therefore need not be solved by the inverse functions in \eqref{eq:extended-lambert-inverse-law}.
The elliptic systems in Section \ref{sec:elliptic-examples} provide immediate examples: their mode equations may be polynomial even though $f(\rho)^2$, rather than $f(\rho)$ itself, is polynomial, and their solutions are generally elliptic functions.

Finally, any such finite rational system may be reduced locally to a polynomial system by a further change of parameter.
Collecting all denominators, write the restricted equations as
\begin{equation}
    \frac{dy_a}{dz}=\frac{A_a(y)}{D(y)},
    \qquad
    y=(u_1,\dots,u_N,v_1,\dots,v_N),
    \qquad a=1,\dots,2N,
    \label{eq:rational-system-common-denominator}
\end{equation}
where the $A_a$ and $D$ are polynomials.
On a region where $D(y)\ne0$, define
\begin{equation}
    \frac{d\xi}{dz}=\frac{1}{D(y(z))}.
\end{equation}
It then follows that
\begin{equation}
    \frac{dy_a}{d\xi}=A_a(y),
\end{equation}
so the rational and polynomial systems trace the same local solution curves with different parametrisations.
This is the same Hamiltonian multiplication and time-reparameterisation mechanism as in Lemma \ref{lem:common-modal-power-factor}, applied in the opposite direction.
Multiplying $H$ by the denominator $D$ gives a Hamiltonian for the reparameterised polynomial equations on $H=0$, since the additional terms produced by the product rule are proportional to $H$ and vanish there.
In particular, using a common modal-power denominator $D_0$ for the rational functions $F$ and $P$ gives a corresponding Kummer-type Hamiltonian with the polynomial functions $D_0F$ and $D_0P$.
An additional factor depending separately on the modes, i.e. not just mode powers, may be needed to clear every denominator in the coupled equations; the resulting Hamiltonian remains canonical on the chosen level set but need not retain the Kummer-type form away from it.

\section{A system generated by the modular \texorpdfstring{$\lambda$}{lambda} function}
\label{sec:modular-example}

We now give a deliberately constructed example generated by the modular lambda function, $\lambda$.
Its first-order differential equation is
\begin{equation}
    \frac{d\lambda}{d\tau}
    =\frac{4i}{\pi}\lambda(1-\lambda)K(\lambda)^2,
    \label{eq:modular-lambda-ode}
\end{equation}
where $K(m)$ denotes the complete elliptic integral of the first kind with parameter $m$.
The familiar theta-constant representation is
\begin{equation}
    \lambda(\tau)
    =\frac{\vartheta_2(\tau)^4}{\vartheta_3(\tau)^4},
    \qquad
    \lambda(\tau)-1
    =-\frac{\vartheta_4(\tau)^4}{\vartheta_3(\tau)^4}.
    \label{eq:modular-lambda-theta}
\end{equation}
Here and below the theta constants have nome $q=e^{i\pi\tau}$.
This example is useful because the generating equation is classical and explicitly solvable, while the resulting functions $P$ and $Q$ are non-polynomial functions of the modal powers.

Take $N=2$ and $\beta_1=\beta_2=2$, and write
\begin{equation}
    x_1=u_1v_1,
    \qquad
    x_2=u_2v_2.
\end{equation}
For parameters $a_0,a_2\in\mathbb{C}$, choose
\begin{align}
    P(x_1,x_2)
    &=\frac{i}{\pi}\sinh(a_2)
    K\left(\frac{1+x_1+x_2}{2}\right)^2,
    \notag\\
    Q(x_1,x_2)
    &=\frac{2i}{\pi}
    K\left(\frac{1+x_1+x_2}{2}\right)^2
    \left[
        \cosh(a_2)x_1x_2
        +a_0(x_1-x_2-1)
    \right].
    \label{eq:modular-lambda-P-Q}
\end{align}
The Hamiltonian \eqref{eq:Hamiltonian} is therefore
\begin{multline}
    H=\frac{i}{\pi}K\left(
        \frac{1+u_1v_1+u_2v_2}{2}
    \right)^2
    \Big[
        \sinh(a_2)(u_1^2u_2^2+v_1^2v_2^2)
        -2\cosh(a_2)u_1u_2v_1v_2
        \\[-0.2\baselineskip]
        -2a_0(u_1v_1-u_2v_2-1)
    \Big].
    \label{eq:modular-lambda-Hamiltonian}
\end{multline}
Apart from the common elliptic-integral factor, the interaction has the quartic form familiar from degenerate four-wave mixing, while the $a_0$ term is linear in the modal powers.

Hamilton's equations for \eqref{eq:modular-lambda-Hamiltonian} initially also contain derivatives of $K$.
On the level set $H=0$, the bracketed expression in \eqref{eq:modular-lambda-Hamiltonian} vanishes, so all terms obtained by differentiating the common elliptic-integral factor vanish.
The equations then reduce to the comparatively simple system
\begin{align}
    \frac{du_1}{d\tau}
    &=\frac{2i}{\pi}
    K\left(\frac{1+u_1v_1+u_2v_2}{2}\right)^2
    \left[
        \sinh(a_2)v_1v_2^2
        -\cosh(a_2)u_1u_2v_2
        -a_0u_1
    \right],
    \notag\\
    \frac{du_2}{d\tau}
    &=\frac{2i}{\pi}
    K\left(\frac{1+u_1v_1+u_2v_2}{2}\right)^2
    \left[
        \sinh(a_2)v_2v_1^2
        -\cosh(a_2)u_2u_1v_1
        +a_0u_2
    \right],
    \notag\\
    \frac{dv_1}{d\tau}
    &=\frac{2i}{\pi}
    K\left(\frac{1+u_1v_1+u_2v_2}{2}\right)^2
    \left[
        -\sinh(a_2)u_1u_2^2
        +\cosh(a_2)v_1u_2v_2
        +a_0v_1
    \right],
    \notag\\
    \frac{dv_2}{d\tau}
    &=\frac{2i}{\pi}
    K\left(\frac{1+u_1v_1+u_2v_2}{2}\right)^2
    \left[
        -\sinh(a_2)u_2u_1^2
        +\cosh(a_2)v_2u_1v_1
        -a_0v_2
    \right].
    \label{eq:modular-lambda-coupled-system}
\end{align}
Thus the non-polynomial dependence appears only through a common function of the two modal powers, while the remaining nonlinearities are cubic in the modes.

The two modal powers obey
\begin{align}
    \frac{d(u_1v_1)}{d\tau}
    &=\frac{d(u_2v_2)}{d\tau}
    \notag\\
    &=-\frac{2i}{\pi}\sinh(a_2)
    K\left(\frac{1+u_1v_1+u_2v_2}{2}\right)^2
    (u_1u_2-v_1v_2)(u_1u_2+v_1v_2),
    \label{eq:modular-lambda-power-evolution}
\end{align}
and hence $x_1-x_2$ is conserved.
In the notation of the general theory,
\begin{equation}
    x_j=\gamma_j-2\rho,
    \qquad
    \gamma_1+\gamma_2=0,
    \qquad
    \rho=-\frac{x_1+x_2}{4}.
\end{equation}
We select the invariant manifold
\begin{equation}
    \gamma_1=\frac{1}{2},
    \qquad
    \gamma_2=-\frac{1}{2},
    \qquad
    x_1-x_2=1.
    \label{eq:modular-lambda-gamma-values}
\end{equation}
On the component of $H=0$ for which
$u_1u_2/(v_1v_2)=\coth(a_2/2)$, the scalar equation is
\begin{equation}
    \rho'
    =\frac{2i}{\pi}
    \left(-2\rho-\frac{1}{2}\right)
    \left(-2\rho+\frac{1}{2}\right)
    K\left(\frac{1}{2}-2\rho\right)^2.
    \label{eq:modular-lambda-rho-ode}
\end{equation}
The change of variable
\begin{equation}
    \rho(\tau)=\frac{1}{4}-\frac{1}{2}\lambda(\tau-\tau_0)
    \label{eq:modular-lambda-rho-solution}
\end{equation}
turns \eqref{eq:modular-lambda-rho-ode} exactly into \eqref{eq:modular-lambda-ode}.
Consequently,
\begin{equation}
    u_1v_1=\lambda(\tau-\tau_0),
    \qquad
    u_2v_2=\lambda(\tau-\tau_0)-1.
    \label{eq:modular-lambda-modal-powers}
\end{equation}
Locally, $\tau_0$ is fixed by the initial modal power through the inverse relation
\begin{equation}
    \tau_0
    =-i\frac{K(1-u_1(0)v_1(0))}{K(u_1(0)v_1(0))},
    \label{eq:modular-lambda-tau0}
\end{equation}
with the modular and elliptic-integral branches chosen consistently.

The individual modes are also explicit.
Integrating the phase equations gives
\begin{align}
    \frac{u_1}{v_1}
    &=\alpha_1
    \left(
        \frac{\lambda(\tau-\tau_0)-1}{\lambda(\tau-\tau_0)}
    \right)^{a_0},
    \notag\\
    \frac{u_2}{v_2}
    &=\alpha_2
    \left(
        \frac{\lambda(\tau-\tau_0)-1}{\lambda(\tau-\tau_0)}
    \right)^{-a_0},
    \label{eq:modular-lambda-mode-ratios}
\end{align}
where $\alpha_1$ and $\alpha_2$ are non-zero constants.
Combining \eqref{eq:modular-lambda-mode-ratios} with \eqref{eq:modular-lambda-modal-powers} yields
the following lambda-function and theta-constant forms.
To keep the display readable, every occurrence of $\lambda$ and $\vartheta_j$ in this equation is evaluated at $\tau-\tau_0$:
\begin{align}
    u_1^2
    &=\alpha_1\lambda
    \left(\frac{\lambda-1}{\lambda}\right)^{a_0}
    =\alpha_1\frac{\vartheta_2^4}{\vartheta_3^4}
    \left(-\frac{\vartheta_4^4}{\vartheta_2^4}\right)^{a_0},
    \notag\\
    u_2^2
    &=\alpha_2(\lambda-1)
    \left(\frac{\lambda-1}{\lambda}\right)^{-a_0}
    =-\alpha_2\frac{\vartheta_4^4}{\vartheta_3^4}
    \left(-\frac{\vartheta_4^4}{\vartheta_2^4}\right)^{-a_0},
    \notag\\
    v_1^2
    &=\frac{\lambda}{\alpha_1}
    \left(\frac{\lambda-1}{\lambda}\right)^{-a_0}
    =\frac{\vartheta_2^4}{\alpha_1\vartheta_3^4}
    \left(-\frac{\vartheta_4^4}{\vartheta_2^4}\right)^{-a_0},
    \notag\\
    v_2^2
    &=\frac{\lambda-1}{\alpha_2}
    \left(\frac{\lambda-1}{\lambda}\right)^{a_0}
    =-\frac{\vartheta_4^4}{\alpha_2\vartheta_3^4}
    \left(-\frac{\vartheta_4^4}{\vartheta_2^4}\right)^{a_0}.
    \label{eq:modular-lambda-mode-solutions}
\end{align}
These expressions are understood locally with consistent logarithmic and square-root branches.
The remaining condition $H=0$ relates $a_2$ to the integration constants:
\begin{equation}
    \tanh(a_2)
    =\frac{2\alpha_1\alpha_2}
    {\alpha_1^2\alpha_2^2+1}.
    \label{eq:modular-lambda-a2-condition}
\end{equation}
Equivalently, away from limiting cases, the two possible branches are
$\alpha_1\alpha_2=\tanh(a_2/2)$ and
$\alpha_1\alpha_2=\coth(a_2/2)$.

The second equality in each line of \eqref{eq:modular-lambda-mode-solutions} follows from \eqref{eq:modular-lambda-theta}.
In particular, when $a_0$ is a half-integer, the square roots can be chosen so that the modes are ratios of integer powers of theta constants and are therefore meromorphic functions of $\tau$.
For example, for $a_0=1/2$ one consistent choice is
\begin{align}
    u_1
    &=e^{i\pi/4}\sqrt{\alpha_1}
    \frac{\vartheta_2\vartheta_4}{\vartheta_3^2},
    &
    u_2
    &=e^{i\pi/4}\sqrt{\alpha_2}
    \frac{\vartheta_2\vartheta_4}{\vartheta_3^2},
    \notag\\
    v_1
    &=e^{-i\pi/4}\frac{1}{\sqrt{\alpha_1}}
    \frac{\vartheta_2^3}{\vartheta_3^2\vartheta_4},
    &
    v_2
    &=-e^{-i\pi/4}\frac{1}{\sqrt{\alpha_2}}
    \frac{\vartheta_4^3}{\vartheta_2\vartheta_3^2},
    \label{eq:modular-lambda-half-integer-solutions}
\end{align}
where every theta constant in \eqref{eq:modular-lambda-half-integer-solutions} is evaluated at $\tau-\tau_0$.

The roles of the parameters are now transparent.
The constant $\tau_0$ translates the modular solution and is fixed by the initial modal powers.
The parameter $a_0$ controls the opposite phase contributions in the two modes but does not alter their powers on the invariant manifold \eqref{eq:modular-lambda-gamma-values}.
The parameter $a_2$ weights the two quartic combinations in the Hamiltonian and, on $H=0$, fixes the product $\alpha_1\alpha_2$ through \eqref{eq:modular-lambda-a2-condition}; the separate values of $\alpha_1$ and $\alpha_2$ determine the distribution between each mode and its Hamiltonian conjugate.

\begin{figure}[H]
    \centering
    \includegraphics[width=0.9\textwidth]{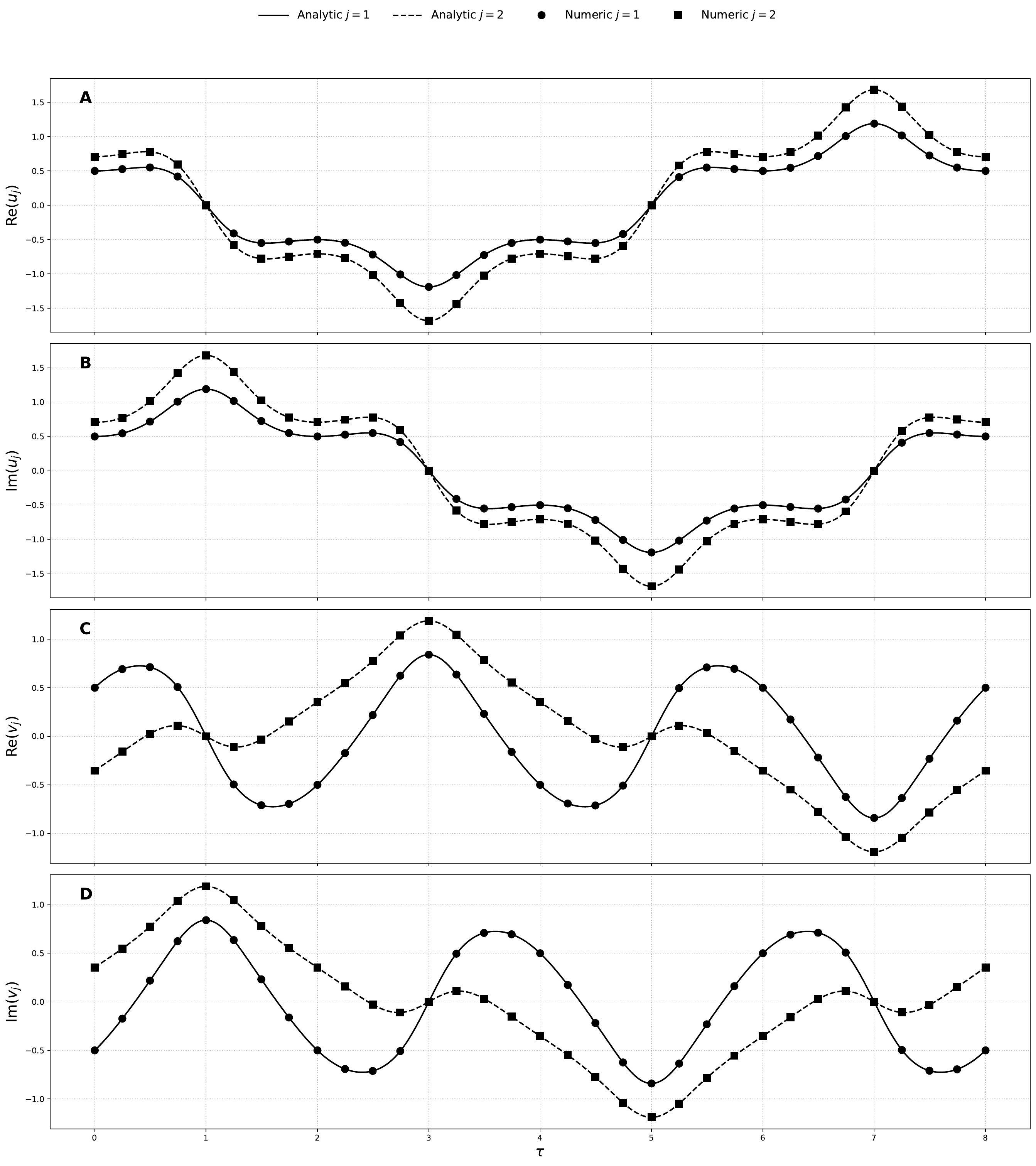}
    \caption{Comparison of the analytic theta-constant solutions (lines) with direct numerical integration of the original Hamilton equations (markers). The parameters are $a_0=1/2$, $a_2=\log 3$, $\alpha_1=1$, $\alpha_2=2$, and $\tau_0=-i$, so that $\alpha_1\alpha_2=\coth(a_2/2)$. Numerical integration uses SciPy \texttt{solve\_ivp} with DOP853, relative tolerance $10^{-10}$ and absolute tolerance $10^{-12}$.}
    \label{fig:modular-lambda-example}
\end{figure}

Figure \ref{fig:modular-lambda-example} compares \eqref{eq:modular-lambda-half-integer-solutions} with an independent numerical integration of the original Hamilton equations, before making the $H=0$ simplification used in \eqref{eq:modular-lambda-coupled-system}.
The elliptic-integral argument in the numerical system is calculated directly from the evolving modal powers, rather than from the modular lambda function.
The numerical values satisfy $\alpha_1\alpha_2=\coth(a_2/2)=2$ and therefore select this same component of $H=0$.

Finally, the construction has a direct relation to the Darboux--Halphen system.
With primes denoting differentiation with respect to $\tau$, we use the convention
\begin{subequations}
\label{eq:Darboux-Halphen-system}
\begin{align}
    h_1'&=h_2h_3-h_1(h_2+h_3),\\
    h_2'&=h_3h_1-h_2(h_3+h_1),\\
    h_3'&=h_1h_2-h_3(h_1+h_2).
\end{align}
\end{subequations}
A standard map from the modular lambda function to a Darboux--Halphen solution is \cite{moralesManifoldWaysDarbouxHalphen2018}
\begin{subequations}
\label{eq:modular-lambda-DH-map}
\begin{align}
    h_1
    &=-\frac{1}{2}\frac{d}{d\tau}
    \log\left(\frac{\lambda'}{\lambda}\right),\\
    h_2
    &=-\frac{1}{2}\frac{d}{d\tau}
    \log\left(\frac{\lambda'}{\lambda-1}\right),\\
    h_3
    &=-\frac{1}{2}\frac{d}{d\tau}
    \log\left(\frac{\lambda'}{\lambda(\lambda-1)}\right).
\end{align}
\end{subequations}
Since \eqref{eq:modular-lambda-modal-powers} recovers both $\lambda$ and $\lambda-1$ directly from the modal powers, every solution of the present system on the selected invariant manifold determines this Darboux--Halphen solution.
More directly, the map may be written entirely in terms of the modal powers and their ratio:
\begin{subequations}
\label{eq:modular-mode-power-DH-map}
\begin{align}
    h_1
    &=-\frac{1}{2}\frac{d}{d\tau}
    \log\left[\frac{d}{d\tau}\log(u_1v_1)\right],\\
    h_2
    &=-\frac{1}{2}\frac{d}{d\tau}
    \log\left[\frac{d}{d\tau}\log(u_2v_2)\right],\\
    h_3
    &=-\frac{1}{2}\frac{d}{d\tau}
    \log\left[
        \frac{d}{d\tau}\log\left(\frac{u_2v_2}{u_1v_1}\right)
    \right].
\end{align}
\end{subequations}
This form holds for every value of $a_0$, since $a_0$ changes the ratios $u_j/v_j$ but not the modal powers $u_jv_j$.
In particular, when $a_0=0$, equation \eqref{eq:modular-lambda-mode-ratios} gives $u_j/v_j=\alpha_j$.
There is then no nontrivial phase evolution and $v_j=\alpha_j^{-1}u_j$, so the distinction between each mode and its Hamiltonian conjugate reduces to a constant rescaling.
The Darboux--Halphen connection is therefore inherited from the generating equation; the focus here is the new canonical two-mode system and its exact theta-constant solutions.

\section{Formulation in terms of total complex modal phase}
\label{sec:mode-power-form}

Throughout this section, all expressions are considered locally with consistent branch choices and where the displayed denominators are nonzero.
We know that via \eqref{eq:uvj-rho-gammaj} we may write the modal power in terms of $\rho(z)$. 
Therefore, if we define a complex modal phase function $\phi_j: \mathbb{C} \to \mathbb{C}$, we can write the modes as:
\begin{align}
    u_j(z) &= \sqrt{\gamma_j - \beta_j\rho(z)}\,e^{\phi_j(z)}, \notag \\
    v_j(z) &= \sqrt{\gamma_j - \beta_j\rho(z)}\,e^{-\phi_j(z)}.
\end{align}
We then define a total complex modal phase $\Phi: \mathbb{C} \to \mathbb{C}$, weighted by degeneracy integers $\beta_m \in \mathbb{Z}\ (m=1,\dots,N)$, as:
\begin{equation}
    \Phi(z) = \sum_{j=1}^{N}\beta_j\phi_j(z)
\end{equation}
From \eqref{eq:-u-v-prod-fs-det} we get:
\begin{align}
    \prod_{j=1}^{N}\left(\frac{u_j}{v_j}\right)^{\beta_j} &= \frac{q + \rho'}{q - \rho'}, \notag \\
    e^{2\Phi} &= \frac{q + f(\rho)}{q - f(\rho)} \label{eq:Phi-q-f-rho}
\end{align}
and thus $\Phi(z)$ can be expressed as a function of $\rho$ and thus as a function of modal powers.
We also get the useful relations:
\begin{align}
    q &= \frac{\rho'}{\tanh{\Phi}}, \notag \\
    p &= \frac{\rho'}{2\sqrt{r}\sinh{\Phi}}, \label{eq:q-p-phi}
\end{align}
or equivalently:
\begin{align}
    \cosh{\Phi} &= \frac{q}{2 \sqrt{r}\,p}, \notag \\
    \sinh{\Phi} &= \frac{\rho'}{2 \sqrt{r}\,p}. \label{eq:cosh-sinh-qpr}
\end{align}
We also have by the definition of $\phi_j$ and $\Phi$ the relations:
\begin{align}
    \cosh{\Phi} &= \frac{\left(\prod_{m=1}^{N} u_{m}^{{\beta}_{m}} + \prod_{m=1}^{N} v_{m}^{{\beta}_{m}}\right)}{2\sqrt{\prod_{j=1}^{N}\left(u_jv_j\right)^{\beta_j}}}, \notag \\
    \sinh{\Phi} &= \frac{\left(\prod_{m=1}^{N} u_{m}^{{\beta}_{m}} - \prod_{m=1}^{N} v_{m}^{{\beta}_{m}}\right)}{2\sqrt{\prod_{j=1}^{N}\left(u_jv_j\right)^{\beta_j}}}. \label{eq:cosh-sinh-u-v}
\end{align}

This viewpoint can be useful when constructing a Hamiltonian system from predefined solutions.
In particular, it enables us to write a Hamiltonian system in which the solutions are related to the predefined solutions by quadrature, in a way that can be thought of as a gauge-like transform.
Let us suppose we start from predefined solutions $u_j, v_j$ which we will refer to hereafter as the prescribed solutions.
Let $\theta_j: \mathbb{C}^N \to \mathbb{C}$ be a function we choose when constructing the prescribed solutions to encode how each of the mode phases depends on all of the mode powers.
The existence of such $\theta_j$ might appear presumptive and restrictive, but on any interval where $\rho$ is locally invertible it is formally possible to write any $z$ dependent phase term as a function of $\rho$ via $z=\rho^{-1}(\rho(z))$, and then replace $\rho$ with a function of modal powers.
Let $\hat{\Phi}: \mathbb{C}^N \to \mathbb{C}$ be the degeneracy weighted sum of mode phases which encodes how the total modal phase depends on individual mode powers. 
Similarly to how $q,p$ relate to $Q,P$ we thus set:
\begin{align}
    \phi_j(z) &= \theta_j(x_1(z), \dots, x_N(z)), \notag \\
    \Phi(z) &= \hat{\Phi}\left(x_1(z), \dots , x_N(z) \right) = \sum_{j=1}^{N}\beta_j\theta_j(x_1(z), \dots, x_N(z)). 
\end{align}
Suppose then we use the functions $\hat{\Phi}$, $\theta_j$, and $F$ to build a Hamiltonian from \eqref{eq:Hamiltonian}, \eqref{eq:f-F}, and \eqref{eq:q-p-phi}, 
such that: 
\begin{align}
    H\left(\bar{u}_1,\dots, \bar{u}_N, \bar{v}_1,\dots, \bar{v}_N\right) =& \left(\prod_{m=1}^{N} \bar{u}_{m}^{{\beta}_{m}} + \prod_{m=1}^{N} \bar{v}_{m}^{{\beta}_{m}}\right) \frac{F{\left(\bar{u}_1 \bar{v}_1, \dots , \bar{u}_N \bar{v}_N \right)}}{2\sqrt{\prod_{j=1}^{N}\left(\bar{u}_j \bar{v}_j\right)^{\beta_j}}\sinh{\hat{\Phi}\left(\bar{u}_1 \bar{v}_1, \dots , \bar{u}_N \bar{v}_N \right)}} \notag \\
    &- \frac{F{\left(\bar{u}_1 \bar{v}_1, \dots , \bar{u}_N \bar{v}_N \right)}}{\tanh{\hat{\Phi}\left(\bar{u}_1 \bar{v}_1, \dots , \bar{u}_N \bar{v}_N \right)}}, \label{eq:H-phihat-F}
\end{align}
where we denote the modes in this Hamiltonian as $\bar{u}_j, \bar{v}_j$ and we will enquire whether $u_j=\bar{u}_j, v_j=\bar{v}_j$.
It is important to stress that \eqref{eq:Phi-q-f-rho} and thus \eqref{eq:q-p-phi} and \eqref{eq:cosh-sinh-qpr} hold on the level set $H=0$.
For clarity, the way to build and use the system in \eqref{eq:H-phihat-F} from prescribed solutions would be as follows:
\begin{itemize}
   \item Decide the autonomous differential equation that gives $\rho'=f\left(\rho\right)=F\left(x_1,\dots,x_N\right)$. Together with the constants $\beta_j,\gamma_j$ and an initial condition, this determines the modal powers $x_j$.
   \item Decide how the mode phases $\phi_j$ are to depend on $x_1,\dots, x_N$ and thus on $\rho$, then sum them weighted by degeneracy to get $\hat{\Phi}\left(x_1,\dots,x_N\right)$ - this now already determines both the solutions and thus the equations of motion in the prescribed system but this may not in general be the Hamiltonian system we are constructing.
   \item Use the constructed $\hat{\Phi}$ function as $\hat{\Phi}\left(\bar{u}_1 \bar{v}_1,\dots, \bar{u}_N \bar{v}_N\right)$ in \eqref{eq:H-phihat-F} but crucially, without using \eqref{eq:Phi-q-f-rho}, \eqref{eq:q-p-phi} and \eqref{eq:cosh-sinh-qpr} to further simplify the Hamiltonian as otherwise it would trivially become zero prior to differentiating.
   \item Apply Hamilton's equations to \eqref{eq:H-phihat-F} to obtain the general equations of motion for $\bar{u}_j, \bar{v}_j$ and only after doing so does one work on the level set $H=0$ and use \eqref{eq:Phi-q-f-rho}, \eqref{eq:q-p-phi} and \eqref{eq:cosh-sinh-qpr} to simplify the equations of motion.
\end{itemize}
Writing $\bar{x}_j=\bar{u}_j\bar{v}_j$, the equations corresponding to \eqref{eq:eoms-rho-only} are
\begin{align}
    \frac{d \log{\bar{u}_j}}{dz} &= -F\partial_{\bar{x}_j}\hat{\Phi} - \frac{\beta_j F}{2 \bar{x}_j}, \notag \\[10pt]
    \frac{d \log{\bar{v}_j}}{dz} &= F\partial_{\bar{x}_j}\hat{\Phi} - \frac{\beta_j F}{2 \bar{x}_j}, \label{eq:eoms-fhat-phihat}
\end{align}
where $F$ and $\hat{\Phi}$ are evaluated at $(\bar{x}_1,\dots,\bar{x}_N)$. These equations imply
\begin{equation}
    \bar{x}_j'=-\beta_jF.
\end{equation}
Choose a reference point $z_0$ at which the Hamiltonian modes and prescribed modes agree:
\begin{equation}
    \bar{u}_j(z_0)=u_j(z_0),
    \qquad
    \bar{v}_j(z_0)=v_j(z_0).
\end{equation}
It follows that $\bar{x}_j(z_0)=x_j(z_0)$. Since the prescribed and Hamiltonian modal powers satisfy the same differential equations with the same initial values, they agree on the corresponding local solution branch:
\begin{equation}
    \bar{x}_j=x_j.
\end{equation}
We henceforth make this identification. The partial derivatives of $\hat{\Phi}$ determine the individual Hamiltonian modal-phase evolution:
\begin{align}
    \frac{1}{2}\frac{d \log{\left(\frac{\bar{u}_j}{\bar{v}_j}\right)}}{dz} &= \bar{\phi}_j' = -F\partial_{x_j}\hat{\Phi} = -F\sum_{k=1}^{N}\beta_k\partial_{x_j}\theta_k. \label{eq:eoms-fhat-phihat-pow-phi}
\end{align}
However, by definition the prescribed individual phase velocity is:
\begin{align}
    \phi_j' = -F\sum_{k=1}^{N}\beta_k\partial_{x_k}\theta_j. \label{eq:prescribed-pow-phi}
\end{align}
Thus, if we define
\begin{equation}
    \xi'_j(z) = \bar{\phi}_j' - \phi_j' = -F\sum_{k=1}^{N}\beta_k\left(\partial_{x_j}\theta_k - \partial_{x_k}\theta_j\right),
\end{equation}
then the constructed Hamiltonian yields the prescribed phase velocities precisely when
\begin{equation}
    \sum_{k=1}^{N}\beta_k\left(\partial_{x_j}\theta_k-\partial_{x_k}\theta_j\right)=0,
    \qquad j=1,\dots,N.
\end{equation}
Otherwise, the solutions of the constructed Hamiltonian system are related to the prescribed solutions by
\begin{align}
    \bar{u}_j&=u_je^{\xi_j}, \notag \\
    \bar{v}_j&=v_je^{-\xi_j}. \label{eq:ubar-vbar-to-u-v-xi-j}
\end{align}
The common initial values at $z_0$ give $\xi_j(z_0)=0$. Let $\eta_j$ express this phase correction as a function of $\rho$, such that
\begin{equation}
    \xi_j(z)=\eta_j(\rho(z)).
\end{equation}
Letting $\rho_0=\rho(z_0)$, we have $\eta_j(\rho_0)=0$, and using $F=\rho'$ the phase correction is given by
\begin{align}
    \frac{d \eta_j}{d \rho} &= -\sum_{k=1}^{N}\beta_k\left(\partial_{x_j}\theta_k - \partial_{x_k}\theta_j\right), \notag \\
    \eta_j(\rho) &= - \int_{\rho_0}^{\rho}
    \sum_{k=1}^{N}\beta_k\left(\partial_{x_j}\theta_k - \partial_{x_k}\theta_j\right)
    ds. \label{eq:eta-j-rho-quadrature}
\end{align}
In the integrand, all partial derivatives are evaluated at
\begin{equation}
    (x_1,\dots,x_N)
    =
    (\gamma_1-\beta_1s,\dots,\gamma_N-\beta_Ns).
\end{equation}
Thus the solutions of the Hamiltonian system are known by quadrature and can be written explicitly in terms of the prescribed modes.
Furthermore, antisymmetry under interchange of $j$ and $k$ gives
\begin{equation}
    \sum_{j=1}^{N}\beta_j\xi'_j
    =
    -F\sum_{j=1}^{N}\sum_{k=1}^{N}\beta_j\beta_k
    \left(\partial_{x_j}\theta_k - \partial_{x_k}\theta_j\right)
    =0.
\end{equation}
Since $\xi_j(z_0)=0$, it follows that
\begin{equation}
    \sum_{j=1}^{N}\beta_j\xi_j=0
\end{equation}
for all $z$. The transformation in \eqref{eq:ubar-vbar-to-u-v-xi-j} therefore preserves each modal power and both product terms in the Hamiltonian:
\begin{align}
    \bar{u}_j\bar{v}_j &= u_jv_j, \notag \\
    \prod_{j=1}^{N}\bar{u}_j^{\beta_j} &= \prod_{j=1}^{N}u_j^{\beta_j}, \notag \\
    \prod_{j=1}^{N}\bar{v}_j^{\beta_j} &= \prod_{j=1}^{N}v_j^{\beta_j}.
\end{align}
Hence the Hamiltonian in \eqref{eq:H-phihat-F} is exactly invariant along the paired solutions, and the transformation in \eqref{eq:ubar-vbar-to-u-v-xi-j} can be regarded as gauge-like.

The following example applies this construction with each prescribed modal phase depending only on its own modal power. This satisfies the phase-compatibility condition identically, so the gauge-like correction vanishes and the prescribed modes are recovered directly.

\subsection{A system generated by an inverse spherical Bessel function}
\label{sec:inverse-spherical-bessel-example}

We now give a deliberately constructed example that combines the inverse procedure of Section \ref{sec:ode-generation} with the total-phase formulation of Section~\ref{sec:mode-power-form}.
Rather than beginning with a familiar Hamiltonian system, we prescribe a locally invertible special function from which we obtain the equation for $\rho$, the mode solutions, and finally a non-polynomial Hamiltonian that generates them.

Let $j_1$ denote the spherical Bessel function of the first kind of order one,
\begin{equation}
    j_1(s)=\frac{\sin s}{s^2}-\frac{\cos s}{s}.
\end{equation}
The real inverse of $j_1$ is multivalued, with locally single-valued branches on intervals between successive zeros of $j_1'$ \cite{stoutemyerInverseSphericalBessel2022}.
Although inverse spherical Bessel functions have themselves been described as generalisations of Lambert $W$, they are different from the inverse laws in Section \ref{sec:rational-generating-equations}.
Writing the sine and cosine in $j_1$ in exponential form introduces both $e^{is}$ and $e^{-is}$, rather than the single exponential-product structure in \eqref{eq:extended-lambert-inverse-law}; correspondingly, the function $1/j_1'(\rho)$ below is not rational in $\rho$.
Choose $c$ such that $j_1'(c)\ne0$ and select the local inverse branch containing $c$ by defining
\begin{equation}
    j_1(\rho(z))=j_1(c)+z,
    \qquad
    \rho(0)=c. \label{eq:inverse-j1-rho-definition}
\end{equation}
Differentiating \eqref{eq:inverse-j1-rho-definition} gives the autonomous equation
\begin{align}
    \rho'
    &=f(\rho)
    =\frac{1}{j_1'(\rho)}, \notag \\
    &=\frac{\rho^3}
    {\rho^2\sin\rho+2\rho\cos\rho-2\sin\rho}. \label{eq:inverse-j1-rho-ode}
\end{align}
The solution is local: it can be continued along the chosen branch only while $j_1'(\rho)\ne0$.

We construct a two-mode system by taking $N=2$ and $\beta_1=\beta_2=1$.
Writing $x_j=u_jv_j$, the intermodal power laws give
\begin{equation}
    x_j=\gamma_j-\rho,
    \qquad
    \gamma_1+\gamma_2=0,
    \qquad
    \rho=-\frac{x_1+x_2}{2}. \label{eq:inverse-j1-modal-powers}
\end{equation}
For constants $a_j,b_j\in\mathbb{C}$, choose the modal phases
\begin{equation}
    \phi_j(x_j)=a_j\operatorname{Si}(b_jx_j),
    \qquad
    \operatorname{Si}(s)=\int_0^s\frac{\sin t}{t}\,dt,
\end{equation}
and hence the total phase
\begin{equation}
    \widehat{\Phi}(x_1,x_2)
    =a_1\operatorname{Si}(b_1x_1)
    +a_2\operatorname{Si}(b_2x_2). \label{eq:inverse-j1-total-phase}
\end{equation}
In the notation of the general construction, $\theta_j(x_1,x_2)=\phi_j(x_j)$, so
\begin{equation}
    \partial_{x_j}\theta_k-\partial_{x_k}\theta_j=0,
    \qquad j,k=1,2.
\end{equation}
The correction in \eqref{eq:eta-j-rho-quadrature} therefore vanishes: $\xi_j'=0$ and, with the matching initial conditions, $\xi_j=0$.
This choice also has the useful derivative
\begin{equation}
    \partial_{x_j}\phi_j
    =a_jb_jj_0(b_jx_j),
\end{equation}
where $j_0(s)=\sin(s)/s$.
The prescribed modes are therefore
\begin{align}
    u_j(z)
    &=\alpha_j\sqrt{\gamma_j-\rho(z)}
    \exp\left[a_j\operatorname{Si}\left(b_j(\gamma_j-\rho(z))\right)\right], \notag \\
    v_j(z)
    &=\frac{1}{\alpha_j}\sqrt{\gamma_j-\rho(z)}
    \exp\left[-a_j\operatorname{Si}\left(b_j(\gamma_j-\rho(z))\right)\right],
    \label{eq:inverse-j1-mode-solutions}
\end{align}
where $\alpha_j\ne0$ are constants satisfying $\alpha_1\alpha_2=1$, as required for the selected $H=0$ branch of the Hamiltonian below, and all square roots and inverse functions are taken on consistent local branches.

To construct the Hamiltonian, define
\begin{align}
    F(x_1,x_2)
    &=f\left(-\frac{x_1+x_2}{2}\right)
    =\frac{1}{j_1'\left(-(x_1+x_2)/2\right)}, \notag \\
    &=\frac{1}{j_1'\left((x_1+x_2)/2\right)}, \label{eq:inverse-j1-F}
\end{align}
where the final equality follows because $j_1$ is odd and $j_1'$ is even.
Applying the construction in \eqref{eq:q-p-phi} gives
\begin{equation}
    P(x_1,x_2)
    =\frac{F(x_1,x_2)}{2\sqrt{x_1x_2}\sinh\widehat{\Phi}(x_1,x_2)},
    \qquad
    Q(x_1,x_2)
    =\frac{F(x_1,x_2)}{\tanh\widehat{\Phi}(x_1,x_2)}. \label{eq:inverse-j1-P-Q}
\end{equation}
Thus a canonical Hamiltonian generating the prescribed power and phase behaviour is
\begin{multline}
    H
    =\frac{F(x_1,x_2)(u_1u_2+v_1v_2)}
    {2\sqrt{x_1x_2}\sinh\widehat{\Phi}(x_1,x_2)}
    -\frac{F(x_1,x_2)}{\tanh\widehat{\Phi}(x_1,x_2)}, \\
    x_1=u_1v_1,
    \qquad
    x_2=u_2v_2. \label{eq:inverse-j1-Hamiltonian}
\end{multline}
Both $P$ and $Q$ are non-polynomial functions of the modal powers.

After applying Hamilton's equations to \eqref{eq:inverse-j1-Hamiltonian} and then working on the level set $H=0$, the resulting coupled system can be written directly in terms of the modes as
\begin{align}
    \frac{du_j}{dz}
    &=-\frac{a_jb_jj_0\left(b_ju_jv_j\right)}
    {j_1'\left((u_1v_1+u_2v_2)/2\right)}u_j
    -\frac{1}
    {2v_jj_1'\left((u_1v_1+u_2v_2)/2\right)}, \notag \\
    \frac{dv_j}{dz}
    &=\frac{a_jb_jj_0\left(b_ju_jv_j\right)}
    {j_1'\left((u_1v_1+u_2v_2)/2\right)}v_j
    -\frac{1}
    {2u_jj_1'\left((u_1v_1+u_2v_2)/2\right)},
    \qquad j=1,2. \label{eq:inverse-j1-coupled-system}
\end{align}
Here $j_1'$ denotes differentiation with respect to the argument of $j_1$.
The first term in each equation is proportional to the corresponding mode and contains $j_0(b_ju_jv_j)$, a function of that mode's own modal power.
It may therefore be interpreted as a generalised self-phase-modulation-like term whose overall strength is modulated by the common total-power factor in the denominator.
The second terms determine the modal-power evolution, and direct substitution into \eqref{eq:inverse-j1-coupled-system} gives
\begin{equation}
    (u_jv_j)'=-\frac{1}{j_1'\left((u_1v_1+u_2v_2)/2\right)}=-\rho',
\end{equation}
in agreement with \eqref{eq:inverse-j1-modal-powers} and \eqref{eq:inverse-j1-rho-ode}.

\begin{figure}[H]
    \centering
    \includegraphics[width=\textwidth]{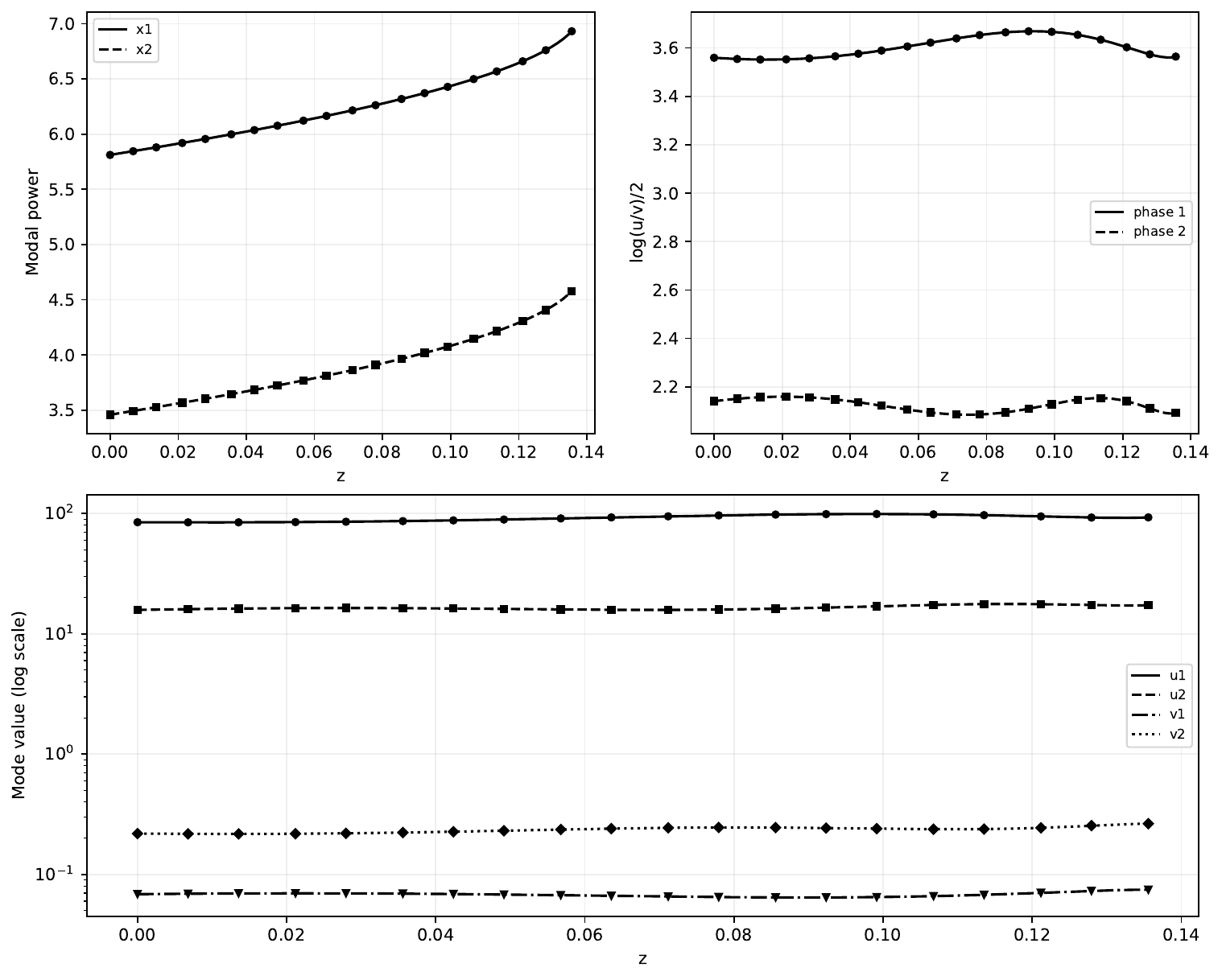}
    \caption{Comparison of the prescribed inverse-spherical-Bessel solutions (lines) with direct numerical integration of the coupled Hamiltonian system (markers). The parameters are $a_1=2.3$, $a_2=1.35$, $b_1=6.4$, and $b_2=9.7$, with $x_1(0)=5.8116$, $x_2(0)=3.4584$, and $\alpha_1=\alpha_2=1$ in \eqref{eq:inverse-j1-mode-solutions}. Thus $u_j(0)=\sqrt{x_j(0)}\exp[a_j\operatorname{Si}(b_jx_j(0))]$ and $v_j(0)=x_j(0)/u_j(0)$, giving $\rho(0)=-4.635$ on the selected branch of $j_1^{-1}$. Numerical integration uses SciPy \texttt{solve\_ivp} with DOP853, relative tolerance $10^{-11}$, absolute tolerance $10^{-13}$, and maximum step $0.05$. The comparison stops at $98\%$ of the interval to the next critical point of $j_1$; the mode panel uses a logarithmic vertical scale.}
    \label{fig:inverse-j1-example}
\end{figure}

Figure \ref{fig:inverse-j1-example} compares the modes in \eqref{eq:inverse-j1-mode-solutions} with a direct numerical integration of the four coupled equations in \eqref{eq:inverse-j1-coupled-system}.
The numerical points lie on the prescribed solutions up to the end of the plotted inverse branch.
The increasingly rapid variation near the right endpoint occurs as $\rho$ approaches the first critical point of $j_1$ on that branch, where $j_1'(\rho)=0$.
This example is intentionally removed from the usual polynomial and elliptic settings: it shows directly how the choice of a locally invertible function can generate a solvable canonical Hamiltonian system involving a different class of special functions.

\section{Discussion}
\label{sec:conclusion}

The construction developed here uses the functional freedom of an established Kummer-type Hamiltonian family to pass from prescribed scalar dynamics to coupled canonical systems.
The intermodal power laws express every modal product as $u_jv_j=\gamma_j-\beta_j\rho$, so a single autonomous equation determines the power evolution and the remaining mode dependence is recovered by quadratures.
Reading this reduction in reverse gives a family of Hamiltonians associated with a chosen scalar equation, or with a locally invertible function defining that equation.
The examples demonstrate how this viewpoint both connects known systems and generates explicit systems across several classes of scalar dynamics.

A scalar equation does not uniquely determine the individual modes.
For a fixed generating equation, the relation between $P$ and $Q$ leaves functional freedom that can change the mode and phase evolution while preserving the prescribed modal powers.
Some apparent differences can be accounted for by transformations already available within the construction.
In particular, Lemma \ref{lem:common-modal-power-factor} shows that, on $H=0$ and where $P\ne0$, the Hamiltonians $H$ and $H/P$ trace the same mode curves with different parametrisations.
This does not remove the freedom associated with prescribing the scalar dynamics in the original variable $z$: the ratio $Q/P$ can still depend on the choice of $P$, and the change of independent variable also depends on that choice.
The total-modal-phase formulation gives a further relation between reconstructed modes through transformations obtained by quadrature.
A natural classification problem is therefore to determine when different choices of Hamiltonian functions produce systems related by canonical transformations, time reparameterisation, or the solution transformations described here, and which features distinguish the remaining choices.

The examples also distinguish the algebraic form of the mode equations from that of their scalar reduction.
Cubic and quartic scalar polynomials give elliptic curves when nonsingular, while higher-degree squarefree polynomials lead to hyperelliptic curves.
Repeated factors require normalization before the genus can be inferred, as the one-mode Fermat family illustrates; its incomplete-beta parametrisation also shows that the curve and the differential defining the independent variable must both be considered.
By contrast, the Lotka--Volterra model and the other log-augmented examples show that polynomial mode equations can retain logarithmic terms in their scalar reductions.
Thus polynomiality of the coupled equations does not by itself imply elliptic or hyperelliptic dynamics.
The examples provide several sufficient mechanisms for rational or polynomial mode equations, even when the Hamiltonian contains nonrational functions. A complete characterisation of such Hamiltonians, with the canonical variables and allowed reparameterisations specified, remains open.

The passage from local quadratures to global mode solutions raises a separate set of questions.
Even when $\rho$ is meromorphic, reconstructing a mode by exponentiating an integral of its logarithmic derivative can introduce branching or nontrivial continuation around closed paths.
For higher-genus scalar curves, the obstruction already occurs in $\rho(z)$ itself: inversion of the scalar Abelian integral generally produces a function with branching under continuation, rather than a globally meromorphic function of $z$. The individual modes may inherit this branching as well as acquiring further branching through reconstruction.
It would be valuable to formulate criteria, in terms of the singularities and periods of the reconstruction differentials, for the modes to be single-valued on a specified domain or meromorphic on an appropriate covering surface.
Such criteria should also clarify continuation through zeros of modal products, where the logarithmic form of the reconstruction equations can become singular even when the original mode equations remain regular.

The non-polynomial examples suggest further constructive questions.
The inverse spherical Bessel example starts directly from a locally invertible special function, while the modular $\lambda$ example yields theta-constant mode solutions and a direct map to the Darboux--Halphen system.
These cases invite a more systematic study of which generating functions permit the reconstruction quadratures to be evaluated within a specified class of special functions.
For algebraic scalar reductions, the determinant identities discussed in the elliptic and hyperelliptic settings suggest investigating how the conserved mode products are encoded by identities on the associated curves.
For nonalgebraic reductions, an analogous description would require structures adapted to the generating functions rather than an assumed algebraic curve.

These questions concern both the scalar equation and its canonical realisations.
The present construction is local and applies on a selected Hamiltonian level and fixed intermodal invariants, with compatible choices of branches.
Extending it therefore involves understanding how those local realisations continue, how they vary with the conserved quantities, and which equivalences preserve their reconstruction data.
This provides a mathematical programme for organising the functional freedom of the Hamiltonian family beyond a list of explicitly solvable examples.

\section*{Code availability}

The computational companion, including self-contained Marimo notebooks, supplementary derivations, executed HTML exports, and symbolic and numerical tests, is available at\newline
\url{https://github.com/HeskethGD/gen-int-sys}.

\appendix

\section{Direct verification of reconstruction}
\label{app:direct-reconstruction}

This appendix proves the reconstruction product identities directly from the differential equations, independently of Hamiltonian conservation. We work in reverse from the equations for $\rho$, $u_j$ and $v_j$ to the identities derived in Section \ref{sec:gen-ham-system}, making the local branch and initial-data compatibility conditions explicit.

\begin{theorem}[Direct reconstruction theorem]
Let $\Omega\subset\mathbb C^N$ and $D\subset\mathbb C$ be open, with $D$ simply connected.
Let $P,Q$ be holomorphic on $\Omega$, and let $u_j,v_j,\rho$ be holomorphic on $D$, with $u_jv_j\ne0$, $x(D)\subset\Omega$, and $P(x(z))\ne0$ throughout $D$.
Let $\beta_j\in\mathbb Z$ satisfy $\sum_j\beta_j\ne0$, and define
\begin{align}
    x_j(z) &= u_j(z) v_j(z), \\
    p(z) &= P\left(x_1,\dots, x_N\right), \\
    q(z) &= Q\left(x_1,\dots, x_N\right), \\
    r(z) &= \prod_{m=1}^{N}x_m^{\beta_m}.
\end{align}
Let prime denote differentiation with respect to $z$, assume $\rho'(z)\ne0$ on $D$, and let $\rho$ satisfy
\begin{equation}
    {\rho'}^2 = q^2 - 4p^2r. \label{eq:theorem-rho-srd}
\end{equation}
Let $u_j$ solve:
\begin{equation}
    \left(\log\left(u_j\right)\right)' = \frac{\beta_j Q}{2 x_j} - \partial_{x_j}Q + \frac{Q \partial_{x_j}P}{P} + \frac{\left(\log\left(x_j\right)\right)'}{2}, \label{eq:theorem-dlog-uj}
\end{equation}
where $P,Q$ and their partial derivatives are evaluated at $(x_1(z),\dots,x_N(z))$. 
Define $v_j$ such that for $\gamma_j \in \mathbb{C} \quad  (j=1,\dots,N): \sum_{j=1}^{N}\gamma_j = 0$:
\begin{equation}
    u_j v_j = \gamma_j - \beta_j \rho. \label{eq:ujvj-rho-theorem}
\end{equation}
Choose $z_0\in D$ and impose the initial compatibility condition
\begin{equation}
    \prod_{m=1}^{N}u_m\left(z_0\right)^{\beta_m} = \frac{q\left(z_0\right) + \rho'\left(z_0\right)}{2 p\left(z_0\right)}. \label{eq:theorem-z0}
\end{equation}
Then it follows that:
\begin{align}
    \prod_{m=1}^{N}u_m^{\beta_m} = \frac{q + \rho'}{2 p}, \notag \\
    \prod_{m=1}^{N}v_m^{\beta_m} = \frac{q - \rho'}{2 p}, \label{eq:u_v-qprho-prods} 
\end{align}

\end{theorem}

\begin{proof}
The hypotheses permit holomorphic logarithms of the nonvanishing factors on $D$. From \eqref{eq:theorem-dlog-uj} we may write:
\begin{equation}
    \frac{1}{2}\sum_{j=1}^{N}\beta_j\left(\log{\left[\frac{u_j}{v_j}\right]}\right)' = \frac{Q}{2}\sum_{j=1}^{N}\frac{\beta_j^2}{x_j} - \sum_{j=1}^{N}\beta_j\partial_{x_j}Q + \frac{Q}{P}\sum_{j=1}^{N}\beta_j\partial_{x_j}P, \label{eq:theorem-log-sum}
\end{equation}
and from \eqref{eq:ujvj-rho-theorem}:
\begin{equation*}
    \beta_j = -\frac{{x_j}'}{\rho'}
\end{equation*}
hence, by making use of general calculus identities such as:
\begin{equation*}
    q' = \sum_{j=1}^{N}{x_j}'\partial_{x_j}Q
\end{equation*}
we may rewrite \eqref{eq:theorem-log-sum} as:
\begin{align}
    \left(\log{\left[\prod_{j=1}^{N}\left(\frac{u_j}{v_j}\right)^{\beta_j}\right]}\right)' &= -\frac{Q}{\rho'}\left(\log{\left[\prod_{j=1}^{N}{x_j}^{\beta_j}\right]}\right)' + \frac{2}{\rho'}\sum_{j=1}^{N}{x_j}'\partial_{x_j}Q - \frac{2Q}{\rho' P}\sum_{j=1}^{N}{x_j}'\partial_{x_j}P, \notag \\[10pt]
    &= -\frac{q r'}{\rho' r} + \frac{2 q'}{\rho'} - \frac{2 q p'}{\rho' p}. \label{eq:theorem-log-sum-2}
\end{align}
However, we also observe that:
\begin{align}
    \left(\log{\left[\frac{q + \rho'}{q - \rho'}\right]}\right)' &= \frac{q' + \rho''}{q + \rho'} - \frac{q' - \rho''}{q - \rho'}, \notag \\[10pt]
    &= -\frac{q r'}{\rho' r} + \frac{2q'}{\rho'} - \frac{2 q p'}{\rho' p}
\end{align}
hence:
\begin{align}
    \left(\log{\left[\prod_{j=1}^{N}\left(\frac{u_j}{v_j}\right)^{\beta_j}\right]}\right)'  &= \left(\log{\left[\frac{q + \rho'}{q - \rho'}\right]}\right)', \notag \\
    \left(\log{\left[\prod_{j=1}^{N}\left(\frac{u_j}{v_j}\right)^{\beta_j}\right]}\right)' + \left(\log{\left[\prod_{j=1}^{N}\left(u_j v_j\right)^{\beta_j}\right]}\right)'  &= \left(\log{\left[\frac{q + \rho'}{q - \rho'}\right]}\right)' + \left(\log{\left[r\right]}\right)', \notag \\
    2\left(\log{\left[\prod_{j=1}^{N}{u_j}^{\beta_j}\right]}\right)'  &= \left(\log{\left[r\left(\frac{q + \rho'}{q - \rho'}\right)\right]}\right)', \notag \\
    2\left(\log{\left[\prod_{j=1}^{N}{u_j}^{\beta_j}\right]}\right)'  &= \left(\log{\left[\frac{\left(q + \rho'\right)^2}{4p^2}\right]}\right)', \notag \\
    \left(\log{\left[\prod_{j=1}^{N}{u_j}^{\beta_j}\right]}\right)'  &= \left(\log{\left[\frac{q + \rho'}{p}\right]}\right)'.
\end{align}
Integrating and accounting for \eqref{eq:theorem-z0} thus gives:
\begin{align}
    \int_{z_0}^{z}\left(\log{\left[\prod_{j=1}^{N}{u_j}^{\beta_j}\right]}\right)'{dz}  &= \int_{z_0}^{z}\left(\log{\left[\frac{q + \rho'}{p}\right]}\right)'{dz}, \notag \\
    \prod_{j=1}^{N}{u_j}^{\beta_j}  &= \frac{q + \rho'}{2p}.
\end{align}
As \eqref{eq:theorem-rho-srd} may be factored into:
\begin{align}
    -\left(\frac{\rho' + q}{2p}\right)\left(\frac{\rho' - q}{2p}\right) = r = \prod_{m=1}^{N}\left(u_m v_m\right)^{\beta_m} = \left(\prod_{m=1}^{N}u_m^{\beta_m}\right)\left(\prod_{m=1}^{N}v_m^{\beta_m}\right),
\end{align}
it follows that:
\begin{equation}
    \prod_{j=1}^{N}{v_j}^{\beta_j}  = \frac{q - \rho'}{2p}.
\end{equation}

\end{proof}

\paragraph{Initial reconstruction data.}
Condition \eqref{eq:theorem-z0} fixes one multiplicative reconstruction constant.
Its right hand side is nonzero because $(q+\rho')(q-\rho')=4p^2r\ne0$.
Since $\sum_j\beta_j\ne0$, at least one weight is nonzero; after choosing the other nonzero initial mode values, that mode can be chosen to satisfy \eqref{eq:theorem-z0}, with a compatible root if necessary, and $v_j(z_0)=x_j(z_0)/u_j(z_0)$.
For nonconstant holomorphic solutions, the product identities extend by continuity across isolated zeros of $\rho'$ wherever the modes and coefficients extend regularly.

\paragraph{Constant modal powers.}
Constant solutions of the squared scalar equation are not covered by the theorem: its derivation divides by $\rho'$, and a scalar root alone need not give a compatible mode solution.
Suppose $\rho(z)\equiv\rho_*$, where $\rho_*$ is a fixed constant, and all modal products are nonzero.
Then $x_j=\gamma_j-\beta_j\rho_*$ and hence $p,q,r$ are constant.
In this case the required product identities reduce to
\begin{equation}
    U(z)=V(z)=\frac{q}{2p},
    \qquad U(z)=\prod_j u_j(z)^{\beta_j},\quad
    V(z)=\prod_j v_j(z)^{\beta_j}.
\end{equation}
Thus the weighted products must be constant in this constant-power case, although they generally vary when $\rho$ is nonconstant.
In addition to the initial product condition and $q^2=4p^2r$, compatibility requires
\begin{equation}
    \sum_{j=1}^{N}\beta_j\left[
    \frac{\beta_j q}{2x_j}-\partial_{x_j}Q
    +\frac{q}{p}\partial_{x_j}P\right]=0,
    \label{eq:stationary-scalar-compatibility}
\end{equation}
with all terms evaluated at $x_j=\gamma_j-\beta_j\rho_*$.
For constant modal powers, the left hand side is $U'/U$. Its vanishing ensures that the initially imposed value $U=q/(2p)$ is preserved; since $UV=r$, the value of $V$ is then preserved as well.
The individual modes may still evolve, with their logarithmic derivatives cancelling in the weighted products.

\begingroup
\emergencystretch=1em
\printbibliography
\endgroup

\end{document}